\documentclass[a4paper,10pt]{article}

\usepackage{authblk}

\usepackage{cmll}
\usepackage{xspace}
\usepackage{multicol}
\usepackage{amsfonts}
\usepackage{amsmath}
\usepackage{amsthm}
\usepackage[disable]{todonotes}
\usepackage{hyperref}
\usepackage{etoolbox}
\usepackage{float}
\floatstyle{boxed} 
\restylefloat{figure}
\usepackage{framed}
\usepackage{txfonts}
\usepackage{graphicx}
\usepackage{caption}
\usepackage{subcaption}
\usepackage{amssymb}
\usepackage{bbold}
\usepackage{bussproofs}
\usepackage{tabularx}

\usepackage{pgf,tikz}
\usepackage{pgfplots}
\usetikzlibrary{matrix,arrows,positioning,backgrounds,decorations,shadows,petri}
\tikzset{
    >=stealth',
    pil/.style={
           ->,
           thick,
           shorten <=2pt,
           shorten >=2pt,}
}

\makeatletter
\newcommand\mmodif[1]{\ifinalign@ #1 \else \ifmmode #1 \else $#1$\xspace \fi \fi} 
\makeatother

\newcommand\at{@} 
\newcommand\+{\mathrm{+}}                
\newcommand\pt{\mathrm{\cdot}}           
\newcommand\dash{\mathrm{\text{-}}}
\newcommand\id{\mmodif{\mathop{id}}}
\newcommand\FV{\mmodif{\mathrm{FV}}}     

\newcommand\rta{\mmodif{\rightarrow}}
\newcommand\Rta{\mmodif{\Rightarrow}}

\newcommand\da{\mathrm{\downarrow}}
\newcommand\Da{\mathrm{\Downarrow^h}}
\newcommand\ua{\mathrm{\uparrow}}
\newcommand\Ua{\mathrm{\Uparrow^h}}
\newcommand\cons{\mathrm{\rta}}

\newcommand\ruledarrow[3]{\stackrel{#1}{\rta}\!\!{}_{#2}^{#3}}
\newcommand\Ruledarrow[3]{\stackrel{#1}{\longrightarrow}\!\!{}_{#2}^{#3}}
\newcommand\St{\mmodif{\Rightarrow_{st}}}

\newcommand\Lra{\Leftrightarrow}

\newcommand\llb{\llbracket}
\newcommand\rrb{\rrbracket}

\newcommand\llc{(\! |}
\newcommand\rrc{|\! )}

\newcommand\I{\mmodif{\boldsymbol{I}}}
\newcommand\J{\mmodif{\boldsymbol{J}}}
\newcommand\Id{\mmodif{\boldsymbol{I}}}

\newcommand\G{\mmodif{\boldsymbol{G}}}

\newcommand\FixP{\mmodif{\boldsymbol{\Theta}}}
\newcommand\Om{\mmodif{\boldsymbol{\Omega}}}
\newcommand\0{\mmodif{\boldsymbol{0}}}
\newcommand\eps{\mmodif{\boldsymbol{\epsilon}}}
\newcommand\bareps{\mmodif{\boldsymbol{\bar\epsilon}}}

\newcommand\Dinf{\mmodif{D_\infty}}
\newcommand\Pinf{\mmodif{P_{\!\infty}}}
\newcommand\Hst{\mmodif{\mathcal{H}^*}}

\newcommand\Nat{\mmodif{\mathbb{N}}}
\newcommand\Lamb{\mmodif{\Lambda}}
\newcommand\Lam[1]{\mmodif{\Lambda_{\tau(#1)}}}
\newcommand\Lcont{\mmodif{\Lambda^{\!\protect\llc\pt\protect\rrc}}}
\newcommand\Lamcont[1]{\mmodif{\Lambda^{\!\protect\llc\pt\protect\rrc}_{\tau(#1)}}}
\newcommand\Test[1]{\mmodif{\boldsymbol{T}_{\tau(#1)}}}
\newcommand\Testc[1]{\mmodif{\boldsymbol{T}^{\protect\llc\pt\protect\rrc}_{\tau(#1)}}}

\newcommand\mhnf{\mathrm{mhnf}}

\newcommand\Achf[1]{\mmodif{\mathcal{A}_f(#1)}}     
\newcommand\Comp[1]{\mmodif{\bar #1}}               
\newcommand\Compl[1]{\mmodif{\overline{#1}}}        

\newcommand\Scottl{\textsc{ScottL}\xspace}  
\newcommand\Scottlb{{\textsc{ScottL}$_!$}\xspace} 

\newcommand\leob{\mmodif{\sqsubseteq_{\mathcal{H}^*}}}
\newcommand\geob{\mmodif{\sqsupseteq_{\mathcal{H}^*}}}
\newcommand\ngeob{\mmodif{\not\sqsupseteq_{\mathcal{H}^*}}}
\newcommand\equivob{\mmodif{\equiv_{\mathcal{H}^*}}}
\newcommand\leobtau[1]{\mmodif{\sqsubseteq_{\tau(#1)}}}

\newcommand\equivobtau[1]{\mmodif{\equiv_{\tau(#1)}}}

\newcommand\Lcalcul{$\lambda$-calculus\xspace}
\newcommand\Lcalculs{$\lambda$-calculi\xspace}

\newcommand\Kweb{K-model\xspace}

\newcommand\Kwebs{K-models\xspace}

\newcommand\newsym[1]{\mmodif{#1}}
\newcommand\newsyminv[1]{}
\newcommand\newsyminvinv[2]{}
\newcommand\newsyminvsec[2]{}
\newcommand\newsymprem[2]{\mmodif{#1}}
\newcommand\newsymsec[2]{\mmodif{#2}}
\newcommand\newsymsecinv[3]{\mmodif{#2}}
\newcommand\newsympreminvinv[3]{\mmodif{#1}}
\newcommand\newsyminvinvinv[3]{}
\newcommand\newdef[1]{{\em #1}}
\newcommand\newdefinv[1]{}
\newcommand\newdefinvinv[2]{}
\newcommand\newdefprem[2]{{\em #1}}
\newcommand\newdefpremsec[2]{{\em #1 #2}}
\newcommand\newdefsecprem[2]{{\em #2 #1}}
\newcommand\newdefsecpreminv[3]{{\em #2 #1}}
\newcommand\newdefinvinvinv[3]{}

\newtheorem{definition}{Definition}
\newtheorem{theorem}[definition]{Theorem}
\newtheorem{example}[definition]{Example}
\newtheorem{lemma}[definition]{Lemma}
\newtheorem{remark}[definition]{Remark}

\newtheorem{proposition}{Proposition}

\newenvironment{oneshot}[1]{\smallskip \noindent{\bf #1} \begin{em}}{\end{em}}
\newenvironment{boite}{\vspace{-5pt}\begin{framed}\begin{center}\vspace{-10pt}}
                       {\end{center}\vspace{-10pt}\end{framed}\vspace{-5pt}}

\newenvironment{grammarLSB}{\begin{tabular}{l l r c l  r}}
                       {\end{tabular}}

\begin{document}

\title{On the characterization of models of \Hst:\\ The operational aspect}

\author[1]{Flavien Breuvart}
\affil[1]{LIPN, UMR 7030, Univ Paris Nord, Sorbonne Paris Cit\'e, France \\
  \texttt{breuvart@pps.univ-paris-diderot.fr}}

\maketitle    

\begin{abstract}
We give a characterization, with respect to a large class of models of untyped $\lambda$-calculus, of those models that are fully abstract for head-normalization, {\em i.e.}, whose equational theory is $\mathcal{H}^*$. An extensional K-model $D$ is fully abstract if and only if it is hyperimmune, {\em i.e.}, non-well founded chains of elements of $D$ cannot be captured by any recursive function.

This article, together with its companion paper~\cite{Bre16SynPart} form the long version of \cite{Bre14}. It is a standalone paper that present a purely syntactical proof of the result as opposed to its companion paper that present an independent and purely semantical proof of the exact same result.
\end{abstract}

\section*{Introduction}
%

The histories of full abstraction and denotational semantics
of \Lcalculs are both rooted in four fundamental articles published in
the course of a year. 

In 1976, Hyland \cite{Hyl75} and Wadsworth \cite{Wad76}
independently\footnote{Notice, however, that the idea already appears
in Wadsworth thesis 3 years earlier.} proved the first full abstraction
result of Scott's $\Dinf$ for \Hst. The following year,
Milner \cite{Mil77} and Plotkin \cite{Plo77} showed respectively \linebreak 
that PCF (a Turing-complete extension of the simply typed \Lcalcul) has a
unique fully abstract model up to isomorphism and that this model is
not in the category of Scott domains and continuous functions.


Later, various articles focused on circumventing Plotkin
counter-example \cite{AJM94,HyOn00} or investigating full abstraction
results for other calculi \cite{AbMc97,lai97,Pao06}. However, hardly anyone
pointed out the fact that Milner's uniqueness theorem is specific to
PCF, while \Hst has various models that are fully abstract but not
isomorphic.

 

The quest for a general characterization of the fully abstract models
of head normalization started by successive refinements of a
sufficient, but unnecessary condition \cite{DFH99,Gou95,Man09},
improving the proof techniques from
1976 \cite{Hyl75,Wad76}. 
x                         
While these results shed some light on various fully abstract
semantics for \Hst, none of them could reach a full characterization.


In this article, we give the first full characterization of the full
abstraction of an observational semantics for a specific (but large)
class of models. 
The class we choose is that of Krivine-models, \linebreak
or K-models \cite{Krivine,Ber00}. 
This class, described in Section~\ref{ssec:K-models}, is essentially
the subclass of Scott complete lattices (or filter
models \cite{CDHL84}) which are prime algebraic. We add two further
conditions: extensionality and test-sensibility. 
Extensionality is a standard and perfectly understood notion that require the model to respect the $\eta$-equivalence, notice that it is a necessary condition for the full abstraction if \Hst. On the other hand, test-sensibility is a new notion that we are introducing but which is equivalent to the more commune notion of approximability (by B/''om trees). Test-sensibility basically states that the model is sensible for an extension of the \Lcalcul called tests.

The extensional and
test-sensible \Kwebs are the objects of our characterization and can be
seen as a natural class of models obtained from models of linear
logic \cite{Gir87}. Indeed, the extensional \Kwebs correspond to the
extensional reflexive objects of the co-Kleisli
category associated with the exponential comonad of Ehrhard's \Scottl
category \cite{Ehr09} (Prop.~\ref{prop:reflOb}).


We achieve the characterization of full abstraction for \Hst in
Theorem~\ref{th:final}: a model $D$ is fully abstract for \Hst iff $D$
is {\em hyperimmune} (Def.~\ref{def:hyperim}). Hyperimmunity is the
key property our study introduces in denotational semantics. This
property is reminiscent of the Post's notion of hyperimmune sets in
recursion theory. Hyperimmunity in recursion theory is not only undecidable, but also
surprisingly high in the hierarchy of undecidable properties (it
cannot be decided by a machine with an oracle deciding the halting
problem) \cite{Nies}.

Roughly speaking, a model $D$ is hyperimmune whenever the
$\lambda$-terms can have access to only well-founded chains of
elements of $D$. In other words, $D$ might have non-well-founded
\linebreak chains $d_0\ge d_1\ge\cdots$, but these chains ``grow'' so fast (for a
suitable notion of growth), that they cannot be contained in the
interpretation of any $\lambda$-term.

The intuition that full abstraction of \Hst is related with a kind of
well-foundation can be found in the literature ({\em e.g.},
Hyland's \cite{Hyl75}, Gouy's \cite{Gou95} or
Manzonetto's \cite{Man09}). Our contribution is to give, with
hyperimmunity, a precise definition of this intuition, at least in the
setting of \Kwebs.

A finer intuition can be described in terms of game
semantics. Informally, a game semantic for the untyped \Lcalcul takes
place in the arena interpreting the recursive type $o = o\rta o$. This
arena is infinitely wide (by developing the left $o$) and infinitely
deep (by developing the right~$o$). Moves therein can thus be
characterized by their nature (question or answer) and by a word over
natural numbers. For example,~$q(2.3.1)$ represents a question in the
underlined~``$o$'' in~$o=o\cons (o\cons o\cons (\underline o \cons
o)\cons o)\cons o$. Plays in this game are potentially infinite
sequences of moves, where a question of the form $q(w)$ is followed by
any number of deeper questions/answers, before an answer $a(w)$ is
eventually provided, if any.

A play like $q(\epsilon),q(1)...a(1),q(2)...a(2),q(3)...$ is
admissible: one player keeps asking questions and is infinitely
delaying the answer to the initial question, but some answers are
given so that the stream is productive. However, the full abstraction
for \Hst forbids non-productive infinite questioning 
like in $q(\epsilon),q(1),q(1.1),q(1.1.1)...$, in general. 
Nevertheless, disallowing {\em all} such strategies is sufficient, but
not necessary to get full abstraction. The hyperimmunity condition is
finer: non productive infinite questioning is allowed {\em as long as}
the function that chooses the next question grows faster than any
recursive function (notice that in the example above that choice is
performed by the constant $(n\mapsto 1)$ function). For example, if
$(u_i)_{i\ge 0}$ grows faster than any recursive function, the play
$q(\epsilon),q(u_1),q(u_1.u_2),q(u_1.u_2.u_3)...$ is perfectly
allowed. 



Incidentally, we obtain a significant corollary (also expressed in
Theorem~\ref{th:final}) stating that full abstraction coincides with
inequational full abstraction for \Hst (equivalence between
observational and denotational orders). This is in contrast to what
happens to other calculi \cite{Sto90,EPT14}. 

In the literature, most of the proofs of full abstraction for \Hst are
based on Nakajima trees \cite{Nak75} or some other notion of quotient
of the space of B\"ohm trees. The usual
approach is too coarse because it considers arbitrary B\"ohm trees
which are not necessarily images of actual $\lambda$-terms. To
overcome this we propose two different techniques leading to two
different proofs of the main result: one purely semantical and the
other purely syntactical. In this article we only present the later, 
the former being the object of a companion paper \cite{Bre16SynPart}.

The semantic proof approaches the problem from a novel angle that 
consists in the use of a new tool: the {\em calculi with tests}
(Def.~\ref{def:LtauD-calculus}). These are syntactic extensions of the
$\lambda$-calculus with operators defining compact elements of the
given models. Since the model appears in the syntax, we are able to
perform inductions (and co-inductions) directly on the reduction steps
of actual terms, rather than on the construction of B\"ohm trees.

The idea of test mechanisms as syntactic extensions of the \Lcalcul
was first used by Bucciarelli {\em et al.} \cite{BCEM11}. Even though
it was mixed  with a resource-sensitive extension, the idea was
already used to define morphisms of the model. Nonetheless, we can
notice that older notions like Wadsworth's labeled
$\lambda\bot$-calculus \cite{Wad76} seem related to calculi with
tests. The calculi with tests are not {\em ad hoc} tricks, but powerful
and general tools.

One of the purposes of this article is to demonstrate the interest of
tests in the study of the relations between denotational and
operational semantics. Calculi with tests are sort of a dual of B\"ohm
trees. While the latter constitutes a syntactical model for
the \Lcalcul; a calculus with tests is a the semantical language for some
K-model. While B\"ohm trees are built upon the \Lcalcul and reduce the
problem of full abstraction to the semantical level; a calculus with
tests is built upon the model and reduces this problem to the
syntactical level. We claim that, regarding relations between
denotational and operational semantics, B\"ohm trees and \Lcalculs
with tests are equally powerful tools, but extend differently to other
frameworks.


\section{Preliminaries and result}
    \label{sec:Theorem}

  \subsection{Preliminaries}
     \label{ssec:preliminaries}
     \subsubsection{Preorders} \label{PrelA}\label{sec:posets} \quad \newline
Given two partially ordered sets $D=(|D|,\le_D)$ and $E=(|E|,\le_E)$, we denote:
\begin{itemize}
  \item $D^{op}=(|D|,\ge_D)$ the reverse-ordered set.
  \item $D\times E=(|D|\times |E|,\le_{D\times E})$ the Cartesian product endowed with the pointwise order: 
    $$(\delta,\epsilon)\le_{D\times E}(\delta',\epsilon')\quad \text{if}\quad \delta\le_D\delta'\quad \text{and}\quad \epsilon\le_E\epsilon'.$$
  \item $\mathcal{A}_f(D)=(|\mathcal{A}_f(D)|,\le_{\mathcal{A}_f(D)})$ the set of finite antichains of $D$ ({\em i.e.}, finite subsets whose elements are pairwise incomparable) endowed with the order :
 $$a\le_{\mathcal{A}_f(D)} b\ \Lra\ \forall \alpha\in a,\exists\beta\in b, \alpha\le_D\beta$$
\end{itemize}
In the following will we use $D$ for $|D|$ when there is no ambiguity. Initial Greek letters $\alpha,\beta,\gamma...$ will vary on elements of ordered sets. Capital initial Latin letters $A,B,C...$ will vary over subsets of ordered sets. And finally, initial Latin letters $a,b,c...$ will denote finite antichains. 

An {\em order isomorphism} between $D$ and $E$ is a bijection $\phi:|D|\rta|E|$ such that $\phi$ and $\phi^{-1}$ are monotone. 

Given a subset $A\subseteq |D|$, we denote $\mathrm{\downarrow} A=\{\alpha\mid\exists\beta\in A,\alpha\mathrm{\le}\beta\}$. We denote by $\mathcal{I}(D)$ the set of {\em initial segments of $D$}, that is  $\mathcal{I}(D)=\{\mathrm{\downarrow}A\mid A\subseteq |D|\}$.
The set $\mathcal{I}(D)$ is a prime algebraic complete lattice with respect to the set-theoretical inclusion. The {\em sups} are given by the unions and the {\em prime elements} are the downward closure of the singletons. The {\em compact elements} are the downward closure of finite antichains. 

The domain of a partial function $f$ is denoted by $Dom(f)$. The {\em graph} of a Scott-continuous function $f:\mathcal{I}(D)\rta \mathcal{I}(E)$ is
\begin{equation} 
  \mathrm{graph}(f)=\{(a,\alpha)\in \mathcal{A}_f(D)^{op}\mathrm{\times}E\mid \alpha\in f(\mathrm{\downarrow} a)\} \label{eq:graphf}
\end{equation}
Notice that elements of $\mathcal{I}(\mathcal{A}_f(D)^{op}\mathrm{\times}E)$ are in one-to-one correspondence with the graphs of Scott-continuous functions from $\mathcal{I}(D)$ to $\mathcal{I}(E)$.

\subsubsection{\Lcalcul} \quad \newline
The $\lambda$-terms are defined up to $\alpha$-equivalence by the following grammar using notation ``{\em \`a la} Barendregt'' \cite{Barendregt} (where variables are denoted by final Latin letters $x,y,z...$): 
\begin{center}
\begin{tabular}{l c c c l}
($\lambda$-terms) & $\Lamb$ & $M,N$ & $::=$ & $x\quad |\quad \lambda x.M\quad |\quad M\ N$
\end{tabular}
\end{center}
We denote by $\FV(M)$ the set of free variables of a $\lambda$-term $M$. Moreover, we abbreviate a nested abstraction $\lambda x_1...x_k.M$ into $\lambda\vec x^{\, k\!} M$, or, when $k$ is irrelevant, into $\lambda\vec x M$. We denote by $M[N/x]$ the capture-free substitution of $x$ by $N$.

\noindent The $\lambda$-terms are subject to the $\beta$-reduction:
\[
 (\beta)\quad\quad (\lambda x.M)\ N\  \ruledarrow{\beta}{}{}\ M[N/x]
\]
A context $C$ is a $\lambda$-term with possibly some occurrences of a hole, {\em i.e.}:
\begin{center}
\begin{tabular}{l c c c l}
  \hspace{-0.2em}(contexts)\hspace{-0.4em} & $\Lamb^{\llc.\rrc}$\hspace{-0.3em} & $C$\hspace{-1em} & $::=$ & $\llc.\rrc\quad |\quad x\quad |\quad \lambda x.C\quad |\quad C_1\ C_2$
\end{tabular}
\end{center}
The writing $C\llc M\rrc$ denotes the term obtained by filling the holes of $C$ by $M$. 
The small step reduction~$\rta$ is the closure of $(\beta)$ by any context, and $\rta_h$ is the closure of $(\beta)$ by the rules:
\begin{center}
  \AxiomC{$M\rta_h M'$}
  \UnaryInfC{$\lambda x.M\rta_h \lambda x.M'$}
  \DisplayProof\hskip 10pt
  \AxiomC{$M\rta_h M'$}
  \AxiomC{\hspace{-3pt}$M$ is an application}
  \BinaryInfC{$M\ N\rta_h M'\ N$}
  \DisplayProof
\end{center}
The transitive reduction $\rta^*$ (resp $\rta_h^*$) is the reflexive transitive closure of $\rta$ (resp $\rta_h$).\\ The big step head reduction, denoted $M\Da N$, is $M\rta_h^*N$ for $N$ in a {\em head-normal form},\linebreak {\em i.e.}, $N=\lambda x_1...x_k.y\ M_1\cdots M_k$, for $M_1,...,M_k$ any terms. We write $M\Da$ for the ({\em head}) {\em convergence}, {\em i.e.}, whenever there is $N$ such that $M\Da N$. 

\begin{example}
  \begin{itemize}
  \item The \newdef{identity term} $\newsym{\protect\I}:=\lambda x.x$ is taking a term and return it as it is:
    $$ \I\ M\quad \rta\quad M.$$
  \item The $n^{th}$ Church numeral, denoted by \newsym{\protect\underline{n}}, and the successor function, denoted by \newsym{\protect\boldsymbol{S}}, are defined by 
    \begin{align*}
      \underline{n} &:= \lambda fx.\underbrace{f\ (f \cdots\ f\ (f}_{n\ \text{times}}\ x)\cdots ), & \boldsymbol{S}&:= \lambda ufx. u\ f\ (f\ x).
    \end{align*}
    Together they provide a suitable encoding for natural numbers, with $\underline n$ representing the $n^{th}$ iteration.
  \item The \newdef{looping term} $\newsym{\protect\Om}:= (\lambda x.xx)\ (\lambda x.xx)$ infinitely reduces into itself, notice that $\Om$ is an example of a diverging term:
    $$ \Om \quad \rta \quad (x\ x)[\lambda y.y\ y/x]\quad = \quad \Om \quad \rta \quad \Om \quad \rta \quad \cdots.$$
  \item The \newdefsecprem{combinator}{Turing fixpoint} $\newsym{\protect\FixP} := (\lambda uv.v\ (u\ u\ v))\ (\lambda uv.v\ (u\ u\ v))$ is a term that computes the least fixpoint of its argument (if it exists):
    \begin{align*}
      \FixP\ M &\rta (\lambda v.v\ ((\lambda uv.v\ (u\ u\ v))\ (\lambda uv.v\ (u\ u\ v)) v))\ M \\
      &= (\lambda v.v\ (\FixP\ v))\ M \\
      &\rta M\ (\FixP\ M).
    \end{align*}
  \end{itemize}
\end{example}

Other notions of convergence exsit (strong, lazy, call by value...), but our study focuses on head convergence, inducing the equational theory denoted by \Hst. 

\begin{definition}\label{observationalPreorder}  
  The {\em observational preorder} and {\em equivalence} denoted $\leob$ and $\equivob$ are given by:
  \begin{align*}
    M &\leob N  & \text{if}\quad & & \forall C ,\ C\llc M\rrc\Da\ \Rta\ C\llc N\rrc\Da,\\
    M&\equivob N & \text{if}\quad & & M\leob N \text{ and } N\leob M.
  \end{align*}
  The resulting (in)equational theory is called \Hst.
\end{definition}

Henceforth, convergence of a $\lambda$-term means head convergence, and full abstraction for \Lcalcul means full abstraction for \Hst.

\begin{definition}
  A model of the untyped \Lcalcul with an interpretation $\llb-\rrb$ is:
  \begin{itemize}
  \item fully abstract (for \Hst) if for all $M, N\in\Lamb$:
    $$M \equivob N \quad \text{if} \quad \llb M\rrb = \llb N\rrb,$$
  \item inequationally fully abstract (for \Hst) if for all $M, N\in\Lamb$:\footnote{It can be generalised by replacing $\subseteq$ by any order on the model.}
    $$M \leob N \quad \text{if} \quad \llb M\rrb \subseteq \llb N\rrb.$$
  \end{itemize}
\end{definition}

Concerning recursive properties of \Lcalcul, we will use the following one:
  \begin{prop}[{\cite[Proposition~8.2.2]{Barendregt} \footnotemark}]\label{prop:autorecursivity}
    \quad\\
    Let $(M_n)_{n\in \Nat}$ be a sequence of terms such that:
    \begin{itemize}
    \item $\forall n\in \Nat, M_n\in \Lamb^0$,
    \item $(n\mapsto M_n)$ is recursive,
    \end{itemize}
    then there exists $F$ such that:
    $$\forall n, F\ \underline{n}\ \rta^* M_n.$$
  \end{prop}
  \footnotetext{This is not the original statement. We remove the dependence on $\vec x$ that is empty in our case and we replace the $\beta$-equivalence by a reduction since the proof of Barendregt \cite{Barendregt} works as well with this refinement.}

  \subsection{K-models} \quad \newline
    \label{ssec:K-models}
%

We introduce here the main semantical object of this article: extensional \Kwebs \cite{Krivine}\cite{Ber00}. This class of models of the untyped \Lcalcul is a subclass of filter models \todo{define filter models?}\cite{CDHL84} containing many extensional models from the continuous semantics, like Scott's $\Dinf$ \cite{Scott}.

\subsubsection{The category \protect\Scottlb} \quad \newline

Extensional \Kwebs 
correspond to the extensional reflexive Scott domains that are prime algebraic complete lattices and whose application embeds prime elements into prime elements \cite{Hut93,Win98}. However we prefer to exhibit \Kwebs as the extensional reflexive objects of the category \Scottlb which is itself the Kleisli category over the linear category \Scottl \cite{Ehr09}.

\begin{definition}
We define the Cartesian closed category \newdef{\protect\Scottlb} \cite{Hut93,Win98,Ehr09}:
\begin{itemize}
\item {\em objects} are partially ordered sets.
\item {\em morphism} from $D$ to $E$ are a Scott-continuous function between the complete lattices $\mathcal{I}(D)$ and $\mathcal{I}(E)$.
\end{itemize}
The {\em Cartesian product} is the disjoint sum of posets. The {\em terminal object} $\top$ is the empty poset. The {\em exponential object} $D\mathrm{\Rta} E$ is $\Achf{D}^{op}\mathrm{\times} E$. Notice that an element of $\mathcal{I}(D\mathrm{\Rta} E)$ is the graph of a morphism from $D$ to $E$ (see Equation \eqref{eq:graphf}). This construction provides a natural isomorphism between $\mathcal{I}(D\mathrm{\Rta} E)$ and the corresponding homset. Notice that if $\simeq$ denotes the isomorphism in \Scottlb, then:
\begin{equation}
  D\Rta D\Rta \cdots \Rta D\simeq (\Achf{D}^{op})^n\times D. \label{eq:defRta}
\end{equation}
For example $D\Rta(D\Rta D)\simeq \Achf{D}^{op}\times(\Achf{D}^{op}\times D)= (\Achf{D}^{op})^2\times D$.
\end{definition}


\begin{remark}\label{rk:equivAchSubset}
  In the literature ({\em e.g.} \cite{Hut93,Win98,Ehr09}), objects are preodered sets and the exponential object $D\Rta D$ is defined by using finite subsets (or multisets) instead of the finite antichains. Our presentation is the quotient of the usual one by the equivalence relation induced by the preorder. The two presentations are equivalent (in terms of equivalence of category) but our choice simplifies the definition of hyperimmunity (Definition~\ref{def:hyperim}).
\end{remark}

\begin{prop} 
  The category \Scottlb is isomorphic to the category prime algebraic complete lattice and Scott-continuous maps.
\end{prop}
\begin{proof}
  Given a poset $D$, the initial segments $\mathcal{I}(D)$ form a prime algebraic complete lattice with $\{\downarrow\alpha\mid \alpha\in D\}$ as prime elements since $I=\bigcup_{\alpha\in I}\downarrow\alpha$. Conversely, the prime elements of a prime algebraic complete lattice form a poset. The two operations are inverse one to the other modulo \Scottlb-isomorphisms or, equivalently, Scott-continuous isomorphisms.
\end{proof}

\subsubsection{An algebraic presentation of K-models}

\begin{definition}[\cite{Krivine}]\label{def:K-model}
  An \newdef{extensional K-model} is a pair $(D,i_D)$ where:
  \begin{itemize}
  \item $D$ is a poset.
  \item \newsym{\protect i_D} is an order isomorphism between $D\mathrm{\Rta} D$ and $D$.
  \end{itemize}
\end{definition}

By abuse of notation we may denote the pair $(D,i_D)$ simply by $D$ when it is clear from the context we are referring to an extensional \Kweb. 



\begin{prop} \label{prop:reflOb}
  Extensional \Kwebs correspond exactly to extensional reflexive objects of \Scottlb, {\em i.e.}, an object $D$ endowed with an isomorphism $abs_D:(D\Rta D)\rta D$  (and $app_D:=abs_D^{-1}$).
\end{prop}
\begin{proof}
  Given a \Kweb $(D,i_D)$, the isomorphism between $D\mathrm{\Rta}D$ and $D$ is given by:
  \begin{align*}
    \forall A &\in \mathcal{I}(D\mathrm{\Rta}D),  &\mathrm{app}_D(A) &= \{i_D(a,\alpha)\mid (a,\alpha)\in A\},\\
    \forall B &\in \mathcal{I}(D), &\mathrm{abs}_D(B) &= \{(a,\alpha)\mid i_D(a,\alpha)\in B\}.
  \end{align*}
  Conversely, consider an extensional reflexive object $(D,app_D,abs_D)$ of \Scottlb. Since $abs_D$ is an isomorphism, it is linear (that is, it preserves all sups). For all $(a,\alpha)\in D\mathrm{\Rta}D$, we have 
  $$\da(a,\alpha)=abs(app(\da(a,\alpha)))=\bigcup_{\beta\in app(\da(a,\alpha))}abs(\da\beta).$$
  Thus there is $\beta\in app(\da(a,\alpha)$ such that $(a,\alpha)\in abs(\da \beta)$, and since $abs(\da \beta)\subseteq\da(a,\alpha)$, this is an equality. Thus there is a unique $\beta$ such that $app_D(a,\alpha)=\da\beta$, this is $i_D(a,\alpha)$.
\end{proof}

In the following we will not distinguish between a \Kweb and its associated reflexive object, this is a model of the pure $\lambda$-calculus.


\begin{definition}\label{def:partialKweb}
An \newdef{extensional partial K-model} is a pair $(E,j_E)$ where $E$ is an object of \Scottlb and $j_E$ is a partial function from $E\mathrm{\Rta}E$ to $E$ that is an order isomorphism between $\mathrm{Dom}(j_E)$ and $E$.
$$ E\quad \stackrel{j_E}{\longleftrightarrow}\quad \mathrm{Dom}(j_E)\quad \subseteq\quad (E\Rta E) $$
\end{definition}

\begin{definition}\label{def:Compl}
  The \newdefpremsec{completion}{of a partial K-model} $(E,j_E)$ is the union 
  $$(\Comp{E},j_{\Comp{E}})=(\bigcup_{n\in\mathbb{N}}E_n,\bigcup_{n\in\mathbb{N}}j_{E_n})$$
  of partial completions $(E_n,j_{E_n})$ that are extensional partial K-models defined by induction on $n$. $(E_0,j_{E_0})=(E,j_E)$ and:
  \begin{itemize}
  \item $|E_{n+1}|\ =\ |E_n|\cup (|E_n\Rta E_n|-Dom(j_{E_n}))$
  \item $j_{E_{n+1}}$ is defined only over $|E_n\Rta E_n|\subseteq |E_{n+1}\Rta E_{n+1}|$ by $j_{E_{n+1}} = j_{E_n}\cup id_{|E_n\Rta E_n|-Dom(j_{E_n})}$
  \item $\le_{E_{n+1}}$ is given by $j_{E_{n+1}}(a,\alpha)\le_{E_{n+1}}(b,\beta)$ if $a\ge_{\Achf{E_n}} b$ and $\alpha\le_{E_n}\beta$.
  \end{itemize}
\end{definition}
Remark that $E_{n+1}$ corresponds to $E_n\Rta E_n$ up to isomorphism, what leads to the equivalent definition:

\begin{prop}
  The completion $(\Comp{E},j_{\Comp{E}})$ of an extensional partial \Kweb $(E,j_E)$ can be described as the categorical $\omega$-colimit (in \Scottl) of $(E'_n)_n$ along the injections $(j_n^{-1})_n$ where $(E'_0,j_0)=(E,j_E)$, $E_{n+1}'=E_n'\Rta E_n'$ and $j^{-1}_{n+1}$ is defined by $j^{-1}_{n+1}(a,\alpha)=(j_{n}(a),j_{n}(\alpha))$ if defined.

  
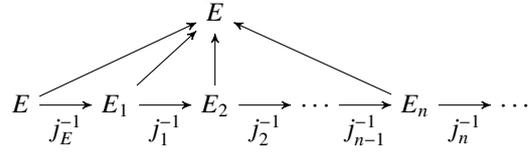
\begin{tikzpicture}[description/.style={fill=white,inner sep=2pt},ampersand replacement=\&]
    \matrix (m) [matrix of math nodes, row sep=2em, column sep=2em, text height=1.5ex, text depth=0.25ex]
    { \& \& \overline{E} \\
      E \& E_1 \& E_2 \& \cdots \& E_n \& \cdots \\
    };
    \path[->] 
      (m-2-1) edge[] node[below] {$j_E^{-1}$} (m-2-2)
      (m-2-2) edge[] node[below] {$j_{1}^{-1}$} (m-2-3)
      (m-2-3) edge[] node[below] {$j_{2}^{-1}$} (m-2-4)
      (m-2-4) edge[] node[below] {$j_{n-1}^{-1}$} (m-2-5)
      (m-2-5) edge[] node[below] {$j_{n}^{-1}$} (m-2-6)
      (m-2-1) edge[] node[auto]  {} (m-1-3)
      (m-2-2) edge[] node[auto]  {} (m-1-3)
      (m-2-3) edge[] node[right] {} (m-1-3)
      (m-2-5) edge[] node[right] {} (m-1-3);
  \end{tikzpicture}
  
\end{prop}
\todo{removable}

\begin{remark}\label{rk:completion}
The completion of an extensional partial \Kweb $(E,j_E)$ is the smallest extensional~\Kweb $\Comp{E}$ containing $E$. In particular, any extensional \Kweb $D$ is the extensional completion of itself: $D=\Comp{D}$.
\end{remark}

\begin{example}\label{example:1}\quad
\begin{enumerate}
  \item \label{enum:Dinf} \newdef{Scott's $\protect\Dinf$} \newsyminvsec{K-model}{\protect\Dinf} \cite{Scott} is the extensional completion of 
    \begin{align*}
      |D| &:= \{*\},   &   \le_D &:= \id,   &   j_D &:= \{(\emptyset,*)\mapsto *\}.
    \end{align*}
    The completion is a triple $(|\Dinf|,\le_{\Dinf},j_{\Dinf})$ where $|\Dinf|$ is generated by:
    \begin{center}
    \begin{grammarLSB}
      & $|\Dinf|$ &   $\alpha,\beta$ & $::=$ & $*\quad |\quad a\cons\alpha$\\
      & $|!\Dinf|$ &  $a,b$ & $\in$ & $\quad \Achf{|\Dinf|}$
    \end{grammarLSB}
    \end{center}
    except that $\emptyset\cons*\not\in|\Dinf|$; $j_{\Dinf}$ is defined by $j_{\Dinf}(\emptyset,*)=*$ and $j_{\Dinf}(a,\alpha)=a\cons\alpha$ \linebreak for $(a,\alpha)\neq(\emptyset,*)$.
  \item \label{enum:Park} \newdef{Park's $\protect\Pinf$} \newsyminvsec{K-model}{\protect\Pinf} \cite{Par76} is the extensional completion of
    \begin{align*}
      |P| &:= \{*\},   &   \le_P &:= \id,   &   j_P &:= \{(\{*\},*)\mapsto *\};
    \end{align*}
    {\em i.e.},  $|P_\infty|$ is defined by the previous grammar except that $(\{*\}\cons *)\not\in |P_\infty|$ while $\emptyset\cons *\in|P_\infty|$. 
  \item \label{enum:D*} \newsymsec{K-model}{Norm} or \newsymsec{K-model}{\protect D^*_\infty} \cite{CDZ87} is the extensional completion of
    \begin{align*}
      |E| &:= \{p,q\},   &   \le_E &:= \id\cup\{p<q\},
    \end{align*}
    \vspace{-2em}
    \begin{align*} 
      j_E &:= \{(\{p\},q)\mathrm{\mapsto} q , (\{q\},p)\mathrm{\mapsto} p\}.
    \end{align*}
  \item \label{enum:WellStrat} \newdef{Well-stratified \Kwebs} \cite{Man09} are the extensional completions of some $E$ respecting 
    $$\forall (a,\alpha)\mathrm{\in} \mathrm{Dom}(j_E), a\mathrm{=}\emptyset.$$
  \item \label{enum:Ind} The \newdef{inductive \Compl{\omega}} is the extensional completion of 
    \begin{align*}
      |E| &:= \mathbb{N},   &   \le_E &:= \id,   &   j_E &:= \{(\{k\mid k< n\},n)\mathrm{\mapsto} n\mid n\in \mathbb{N}\}.
    \end{align*}
  \item \label{enum:CoInd} The \newdef{co-inductive \Compl{\mathbb{Z}}} is the extensional completion of 
    \begin{align*}
      |E| &:= \mathbb{Z},   &   \le_E &:= \id   &,   j_E &:= \{(\{n\},n+1)\mathrm{\mapsto} n\+1\mid n\in \mathbb{Z}\}.
    \end{align*}
  \item \label{enum:Func} \newdef{Functionals $H^f$} \newsyminvsec{K-model}{\protect H^f} (given $f:\Nat\rta\Nat$) are the extensional completions of:
    \begin{align*}
      |E|  &:= \{*\}\cup\{\alpha_j^n \mid n\ge 0,\ 1\le j\le f(n)\},   &
      \le_E &:= \id,
    \end{align*}
    \vspace{-1em}
    \[
      \quad\quad j_E := \ \Big\{(\emptyset, *)\mapsto*\Big\}\ \cup\ \Big\{(\emptyset, \alpha_{j+1}^n)\mapsto\alpha_j^n\mid 1\le j< f(n)\Big\}\ \cup\ \Big\{(\{\alpha_{1}^{n+1}\},*)\mapsto\alpha_{f(n)}^n\mid n\in\Nat^*\Big\},
    \]
    where $(\alpha^n_j)_{n,j}$ is a family of atoms different from $*$.
  \end{enumerate}
\end{example}

\medskip

\noindent For the sake of simplicity, from now on we will work with a fixed extensional \Kweb $D$. Moreover, we will use the notation $a\cons\alpha:=i_D(a,\alpha)$ \newsyminvsec{reduction}{\protect\rta}. Notice that, due to the injectivity of $i_D$, any $\alpha\in D$ can be uniquely rewritten into $a\cons\alpha'$, and more generally into $a_1\cons\cdots\cons a_n\cons\alpha_n$ for any $n$. 

\begin{remark}
  Using this notations, the model $H^f$ can be summarized by writing, for each $n$:
  \[ \alpha_1^n\ =\ \underbrace{\emptyset\cons \cdots \cons \emptyset}_{f(n)}\cons \{\alpha_1^{n\+1}\}\cons * \]
\end{remark}

\subsubsection{Interpretation of the \Lcalcul} \quad \newline

The Cartesian closed structure of \Scottlb endowed with the isomorphisms $app_{D}$ and $abs_{D}$ of the reflexive object induced by $D$ (see Proposition~\ref{prop:reflOb}) defines a standard model of the \Lcalcul.\todo{REF!}

A term $M$ with at most $n$ free variables $x_1,\dots,x_n$ is interpreted as the graph of a mor-\linebreak phism~$\llb M\rrb^{x_1...x_n}_{D}$ \newsyminvinvinv{interpretation}{\protect\llb .\protect\rrb^{\protect\vec x}_{D}}{\protect\Lamb in a K-model} from $D^n$ to $D$ (when $n$ is obvious, we can use \newsymsecinv{interpretation}{\protect\llb.\protect\rrb^{\protect\bar x}}{$\Lambda$ in a K-model}). By Equations \eqref{eq:graphf} and \eqref{eq:defRta} we have:
$$\llb M\rrb^{x_1...x_n}_{D}\subseteq\  (D\Rta\!\cdots\Rta D\Rta D)\ \simeq\ (\Achf{D}^{op})^n\times D. $$
In Figure~\ref{fig:intLam}, we explicit the interpretation $\llb M\rrb_{D}^{x_1...x_n}$ by structural induction on $M$.
\begin{figure*} \label{fig:TestInter}
  \caption{Direct interpretation of $\Lamb$ in $D$ \label{fig:intLam}}
    \begin{center}
      $\llb x_i \rrb_D^{\vec x}  =  \{(\vec a,\alpha)\ |\ \alpha\le\beta\in a_i\}$
      \hspace{6em}
      $ \llb \lambda y.M \rrb_D^{\vec x}  =  \{(\vec a,b\cons\alpha)\ |\ (\vec ab,\alpha)\in\llb M\rrb_D^{\vec xy}\}$
      \vspace{0.3em}\\
      $\!\llb M\ N \rrb_D^{\vec x}  =  \{(\vec a,\alpha)\ |\ \exists b,(\vec a,b\cons\alpha)\in\llb M\rrb_D^{\vec x}\ \wedge \forall \beta\mathrm{\in} b, (\vec a,\beta)\in\llb N\rrb_D^{\vec x}\}$
    \end{center}
\end{figure*}

\begin{example}
  \vspace{-0.5em}
  \begin{align*}
    \llb\lambda x.y\rrb_{D}^y &= \{((a), b\cons\alpha)\mid \alpha\le_{D}\beta\in a\},\\
    \llb\lambda x.x\rrb_{D}^y &= \{((a), b\cons\alpha)\mid \alpha\le_{D}\beta\in b\},\\
    \llb\I\rrb_{D} &= \{a\cons\alpha\mid \alpha\le_{D}\beta\in a\},\\
    \llb \underline{1}\rrb_{D} &= \{a\cons b\cons\alpha \mid \exists c, c\cons\alpha\le_{D}\beta\in a,\ c\le_{\Achf{D}}b\}.
  \end{align*}
  In the last two cases, terms are interpreted in an empty environment. We, then, omit the empty sequence associated with the empty environment, {\em e.g.}, $a\cons b\cons\alpha$ stands for $((),a\cons b\cons\alpha)$.\\
  We can verify that extensionality holds, indeed $\llb \underline{1}\rrb_{D}=\llb\I\rrb_{D}$, since $c\cons\alpha\le_{D}\beta\in a$ and $c\le_{\Achf{D}}b$ exactly say that $b\cons\alpha\!\le_{D}\!\beta\!\in\! a$, and since any element of $\gamma\!\in\! D$ is equal to $d\cons\delta$ for a suitable $d$ and $\delta$.
\end{example}

\subsubsection{Intersection types} \quad \newline

It is folklore that the interpretation of the \Lcalcul into a given \Kweb $D$ is characterized by a specific \newdefprem{intersection type system}{characterizing K-models}. In fact any element $\alpha\in D$ can be seen as an intersection type 
\begin{align*}
  \alpha_1\wedge\cdots\wedge\alpha_n&\rta\beta   &   \text{given by}\quad &\alpha= \{\alpha_1,\dots,\alpha_n\}\cons\beta.
\end{align*}
In Figure~\ref{fig:tyLam}, we give the intersection-type assignment corresponding to the \Kweb induced by $D$.

\begin{figure*}
  \caption{Intersection type system computing the interpretation in $D$ \label{fig:tyLam}}
  \begin{center}
    \AxiomC{$\alpha\in a$}
    \UnaryInfC{$x:a\vdash x:\alpha$}
    \DisplayProof\hskip 50pt
    \AxiomC{$\Gamma\vdash M:\alpha$}
    \UnaryInfC{$\Gamma,x:a\vdash M:\alpha$}
    \DisplayProof\hskip 50pt
    \AxiomC{$\Gamma\vdash M:\beta$}
    \AxiomC{$\alpha\le\beta$}
    \BinaryInfC{$\Gamma\vdash M:\alpha$}
    \DisplayProof\\ \vspace{1em}
    \AxiomC{$\Gamma,x:a\vdash M:\alpha$}
    \UnaryInfC{$\Gamma\vdash \lambda x.M:a\cons\alpha$}
    \DisplayProof\hskip 50pt
    \AxiomC{$\Gamma\vdash M:a\cons\alpha$}
    \AxiomC{$\forall \beta\in a,\ \Gamma\vdash N:\beta$}
    \BinaryInfC{$\Gamma\vdash M\ N:\alpha$}
    \DisplayProof
  \end{center}
\end{figure*}

\begin{prop}
  Let $M$ be a term of \Lamb, the following statements are equivalent:
  \begin{itemize}
  \item $(\vec a,\alpha)\in\llb M\rrb_D^{\vec x}$,
  \item the type judgment $\vec x:\vec a\vdash M:\alpha$ is derivable by the rules of Figure~\ref{fig:tyLam}.
  \end{itemize}
\end{prop}
\begin{proof}
By structural induction on the grammar of $\Lamb$.
\end{proof}

  \subsection{The result} \quad \newline
    \label{ssec:results}
%


We state our main result, showing an equivalence between hyperimmunity (Def.~\ref{def:hyperim}) and full abstraction for \Hst. 


\begin{definition}[\newdef{Hyperimmunity}]\label{def:hyperim}
  A (possibly partial) extensional \Kweb $D$ is said to be \newdef{hyperimmune} if for every sequence  $(\alpha_n)_{n\ge 0}\in D^\Nat$, there is no recursive function $g:\Nat\mathrm{\rta}\Nat$ satisfying, the following condition for all $n\mathrm{\ge} 0$:
  \begin{align}
    \alpha_n&=a_{n,1}\cons\cdots\cons a_{n,g(n)}\cons\alpha_n' & \text{and} & & \alpha_{n\+1}&\in \bigcup_{k\le g(n)}a_{n,k}. \label{eq:hyperimmunity}
  \end{align}
\end{definition}

Notice, in the above definition, that each antichain $a_{n,i}$ always exist and are uniquely determined by the isomorphism between $D$ and $D\Rta D$ that allow us to unfold any element $\alpha_i$ as an arrow (of any length).

The idea is the following. The sequence $(\alpha_n)_{n\ge 0}$ is morally describing a non well-founded chain of elements of $D$, through the isomorphism $D\simeq D\Rta D$, allowing us to see any element $\alpha_i$ as an arrow (of any length):
\begin{alignat*}{4}
  \alpha_0= a_{0,1}\cons \cdots\; &  a_{0,i_0}\cdots \cons a_{0,g(0)}\cons \alpha_0' \hspace{-17pt}\\
&\ \rotatebox[origin=c]{90}{$\in$}\\
\vspace{-1em}
& \alpha_1= a_{1,1}\cons \cdots\; & a_{1,i_1}\cdots \cons a_{1,g(1)}\cons \alpha_1' \hspace{-21pt}\\
& & \omit\span\ \rotatebox[origin=c]{90}{$\in$}\\
& &\alpha_2=  a_{2,1}\cons \cdots &\ a_{2,i_2} \cdots \cons\; a_{2,g(2)}\cons \alpha_2'\\
& & & \quad \ddots
\end{alignat*}
The growth rate $(i_n)_n$ of the chain $(\alpha_n)_n$ depends on how many arrows must be displayed in $\alpha_i$ in order to see $\alpha_{i+1}$ as an element of the antecedent of one of them. Now, hyperimmunity means that if any such non-well founded chain $(\alpha_n)_n$ exists, then its growth rate $(i_n)_n$ cannot be bounded by any recursive function $g$.

\begin{remark}
  It would not be sufficient to simply consider the function $n\mapsto i_n$ such that $\alpha_{n+1}\mathrm{\in} a_{n,i_n}$ rather than the bounding function $g$. Indeed, $n\mapsto i_n$ may not be recursive even while $g$ is.
\end{remark}


\begin{prop} \label{prop:hypPartial}
For any extensional partial \Kweb $E$ (Def.~\ref{def:partialKweb}), the completion $\Compl{E}$ (Def.~\ref{def:Compl}) is hyperimmune iff $E$ is hyperimmune.
\end{prop}
\begin{proof}
  The left-to-right implication is trivial.\\
  The right-to-left one is obtained by contradiction:\\ 
  Assume to have a $(\alpha_n)_{n\ge 0}\in \Comp{E}^{\Nat}$ and a recursive function $g:\Nat\rta\Nat$ such that for all $n\ge 0$:
  \begin{align*}
    \alpha_n&=a_{n,1}\cons\cdots\cons a_{n,g(n)}\cons \alpha_n'  &  \text{and} & & \alpha_{n+1}\in\bigcup_{i\le g(n)} a_{n,i}
  \end{align*}
  Recall that the sequence $(E_k)_{k\ge 0}$ of Definition \ref{def:Compl} approximates the completion $\Comp{E}$.\\
  Then we have the following:
  \begin{itemize}
    \item There exists $k$ such that $\alpha_0\in E_k$, because $\alpha_0\in\Comp{E}=\bigcup_kE_k$.
    \item If $\alpha_n\in E_{j+1}$, then $\alpha_{n+1}\in E_{j}$, because there is $i\le g(n)$ such that $\alpha_{n+1}\in a_{n,i}\subseteq E_j$.
    \item If $\alpha_n\in E_0=E$, then $\alpha_{n+1}\in E$ by surjectivity of $j_E$.
    \end{itemize}
  Thus there is $k$ such that $(\alpha_n)_{n\ge k}\in E^\Nat$, which would break hyperimmunity of $E$.
\end{proof}

\begin{example}\label{ex:hyperimmunity}
  \begin{itemize}
  \item The well-stratified \Kwebs of Example~\ref{example:1}\eqref{enum:WellStrat} (and in particular $\Dinf$ of Item~\eqref{enum:Dinf}) are trivially hyperimmune: already in the partial K-model, there are not even $\alpha_1$, $\alpha_2$ and $n$ such that $\alpha_1=a_{1}\cons\cdots\cons a_{n}\cons\alpha_1'$ and $\alpha_2\in a_n$ (since $a_n=\emptyset$). The non-hyperimmunity of the partial \Kweb can be extended to the completion using Proposition~\ref{prop:hypPartial}.
  \item The model $\Compl{\omega}$ (Ex.~\ref{example:1}\eqref{enum:Ind}) is hyperimmune. Indeed, any such $(\alpha_n)_n$ in the partial K-model would respect $\alpha_{n+1}\mathrm{<_\Nat} \alpha_n$,hence $(\alpha_n)_n$ must be finite by well-foundation of \Nat.
  \item The models \Pinf, $\Dinf^*$ and $\Compl{\mathbb{Z}}$ (Examples~\ref{example:1}\eqref{enum:Park}, \eqref{enum:D*} and \eqref{enum:CoInd}) are not hyperimmune. Indeed for all of them $g=(n\mapsto 1)$ satisfies the condition of Equation \eqref{eq:hyperimmunity}, the respective non-well founded chains $(\alpha_i)_i$ being $(*,*,\dots)$, $(p,q,p,q,\dots)$, and $(0,-1,-2,\dots)$:
      \begin{alignat*}9
        *=\ & \{*\} \rta *\hspace{-15pt} & & & 
        p=\ & \{q\} \rta p\hspace{-15pt} & & & 
        0=\ & \{1\} \rta 0 \hspace{-20pt}\\
        &\ \rotatebox[origin=c]{90}{$\in$} & & & 
        &\ \rotatebox[origin=c]{90}{$\in$} & & &
        &\ \rotatebox[origin=c]{90}{$\in$}\\
        &\ \ *=\  & \omit\span \{*\}\rta *  & 
        &\ q=\ & \omit\span \{p\}\rta q &
        &\ 1=\ &\omit\span \{2\}\rta 1\\
        & &\omit\span\ \ \rotatebox[origin=c]{90}{$\in$} &
        & &\omit\span\ \ \rotatebox[origin=c]{90}{$\in$} & 
        & &\omit\span\ \ \rotatebox[origin=c]{90}{$\in$}\\
        & &\ \ *=\ & \{*\}\rta * &
        & &\omit\span\ p= \{q\}\rta p &
        & &\omit\span\ 2= \{3\}\rta 2\\
        & & &\quad \ddots  & 
        & & \quad\quad\quad \ddots & &
        & & \quad\quad\quad \ddots
      \end{alignat*}
    \item More interestingly, the model $H^f$ (Ex.~\ref{example:1}\eqref{enum:Func}) is hyperimmune iff $f$ is a \emph{hyperimmune function} \cite{Nies}, {\em i.e.}, iff there is no recursive $g:\Nat\rta\Nat$ such that $f\le g$ (pointwise order); otherwise the corresponding sequence is $(\alpha_1^i)_i$.
    \begin{alignat*}4
      \alpha_1^0= \underbrace{\emptyset \rta \cdots \rta \emptyset}_{_{f(0)\ \text{times}}} \rta & \{\alpha_1^1\} \rta  \emptyset \rta \cdots \rta \emptyset\rta * \hspace{-20pt} \\ 
      &\ \rotatebox[origin=c]{90}{$\in$}\\
      & \ \alpha_1^1= \underbrace{\emptyset \rta \cdots \rta \emptyset}_{_{f(1)\ \text{times}}} \rta & \omit\span  \{\alpha_1^2\} \rta  \emptyset \rta \cdots \rta \emptyset \rta *\\
      & & \rotatebox[origin=c]{90}{$\in$} \hspace{7.5em}\\
      & & \ \alpha_1^2= \underbrace{\emptyset \rta \cdots\rta \emptyset}_{_{f(2)\ \text{times}}} \rta & \{\alpha_1^3\} \rta  \emptyset \rta \cdots \rta \emptyset \rta  *\\
      & & &\ \ \rotatebox[origin=c]{90}{$\in$}\\
      & & &\quad  \ddots\\
    \end{alignat*}
  \end{itemize}
\end{example}

The following theorem constitutes the main result of the paper. It shows the equivalence between hyperimmunity and (inequational) full abstraction for \Hst under a certain condition. This conditions, namely the test-sensibility, is a new property that will be defined in more details in Definition~\ref{def:sensibilityWTests}.

\begin{theorem}\label{th:final}
  For any extensional and test-sensible (Def.~\ref{def:sensibilityWTests}) \Kweb $D$, the following are equivalent:
  \begin{enumerate}
  \item $D$ is hyperimmune, \label{eq:n1}
  \item $D$ is inequationally fully abstract for \Hst, \label{eq:n2}
  \item $D$ is fully abstract for \Hst. \label{eq:n3}
  \end{enumerate}
\end{theorem}

\smallskip

\begin{example}
  The model \Dinf (Ex.\ref{example:1}\eqref{enum:Dinf}), the model \Compl{\omega} (Ex.\ref{example:1}\eqref{enum:Ind}) and the well-stratified \Kwebs (Ex.\ref{example:1}\eqref{enum:WellStrat}) will be shown inequationally fully abstract, as well as the models~$H^f$ when~$f$ is hyperimmune. The models~$\Dinf^*$,~\Compl{\mathbb{Z}} (Ex.\ref{example:1}\eqref{enum:D*} and Ex.\ref{example:1}\eqref{enum:CoInd}) will not be, as well as the model~$H^f$ for~$f$ not hyperimmune.
\end{example}

%
%

    \label{sec:proof2}
%

As for the traditional proof of full abstraction for the \Hst, the main idea of our proof is to use a middle step between our calculus and our models. However, this time the proxy will not be a kind of syntactical model (the B\"ohm trees), but a kind of semantical calculus, more exactly a set of calculi that we call \Lcalculs with $D$-tests (Def.~\ref{def:LtauD-calculus}). The traditional interest over B\"ohm trees lies in the fact that they are ``syntactical models'' directly inspired by the calculus (here the \Lcalcul); thus, taking the opposite view, we will use ``semantical calculi'' that are directly inspired by the model (and that are dependent on the \Kweb $D$).

Given a \Kweb $D$, the \Lcalcul with $D$-tests, denoted \Lam{D}, is an extension of the untyped~\Lcalcul that can itself be interpreted in $D$ (Def.~\ref{def:LtauD-calculus}):
\begin{center}
  \begin{tikzpicture}[description/.style={fill=white,inner sep=2pt},ampersand replacement=\&]
    \matrix (m) [matrix of math nodes, row sep=1em, column sep=5em, text height=1.5ex, text depth=0.25ex]
    { {\Lamb} \& \& D \\
      \& {\Lam{D}} \\};
    \path[->] (m-1-1) edge[] node[description] {$\llb.\rrb$} (m-1-3);
    \path[->] (m-1-1) edge[] node[description] {$\subseteq$} (m-2-2);
    \path[->] (m-2-2) edge[] node[description] {$\llb.\rrb$} (m-1-3);
  \end{tikzpicture}
\end{center}

The interest of \Lam{D} relies on the definition of sensibility for \Lam{D} (Def.~\ref{def:sensibilityWTests}), which easily implies the full abstraction of $D$ for \Lam{D} (Th.~\ref{th:FAwT}), even if not for the \Lcalcul. Therefore, it remains to understand when the observational equivalence is preserved from~\Lamb to~\Lam{D}:
\begin{center}
  \begin{tikzpicture}[description/.style={fill=white,inner sep=2pt},ampersand replacement=\&]
    \matrix (m) [matrix of math nodes, row sep=1em, column sep=5em, text height=1.5ex, text depth=0.25ex]
    { {\Lamb} \& {\Lam{D}} \\
      M \&  M \\
      \\
      N \&  N \\
    };
    \path[->] (m-1-1) edge[] node[description] {$\subseteq$} (m-1-2);
    \path[|->] (m-2-1) edge[] node[description] {$id$} (m-2-2);
    \path[-] (m-2-1) edge[] node[description] {$\equivob$} (m-4-1);
    \path[-] (m-2-2) edge[] node[description] {$\quad \equivobtau{D}$} (m-4-2);
    \path[|->] (m-4-1) edge[] node[description] {$id$} (m-4-2);
  \end{tikzpicture}
\end{center}

The proof splits in the two directions: inequational full abstraction implies hyperimmunity (Sec.~\ref{ssec:HtoFA2} and Th.~\ref{th:FA}) and the non-full abstraction for \Hst gives a counterexample to hyperimmunity (Sec.~\ref{ssec:FAtoH2} and Th.~\ref{th:countex2}). However, the proofs will rely on syntactical properties of \Lam{D} such as confluence (Th.~\ref{th:confluence}) and standardization (Th.~\ref{th:standardisation}). 


\section{$\lambda$-calculi with D-tests}
    \label{ssec:D-tests}
%

\subsection{Syntax}\label{sec:TestSynatax}\ \newline
The original idea of using {\em tests} to recover full abstraction (via a theorem of definability) is due to Bucciarelli {\em et al.} \cite{BCEM11}. Here we define variants of Bucciarelli {\em et al.}'s calculus adapted to our framework.

Directly dependent on a given \Kweb $D$, the \Lcalcul with $D$-tests $\Lam{D}$ is, to some extent, an internal calculus for $D$. In fact, we will see that, for $D$ to be fully abstract for \Lam{D}, it is sufficient to be sensible (Th.~\ref{th:FAwT}).

The idea is to introduce tests as a new kind in the syntax. Tests \newdefinv{tests} $Q\in\newsym{\protect\Test{D}}$ are sort of co-terms, in the sense that their interpretations are maps from the context to the dualizing object of the linear category \Scottl ($\bot=\{*\}$):
   $$\llb Q\rrb^{x_1...x_n}\in D^n\Rta \bot$$
The type $\bot$ is the unit type, having only one value representing the convergence of the evaluation, seen as a success.\footnote{We will see in Remark~\ref{rk:polarisedTests} that in a polarized context, the behavior of test does not correspond to co-term (or stack), but to commands (or processes), {\em i.e.}, to interactions between usual terms and fictive co-terms extracted from the semantics.}

The interaction between terms and tests is carried out by two groups of operations indexed by the elements $\alpha\in D$:
  $$\tau_\alpha:\Lam{D}\rta\Test{D}\quad \text{ and }\quad \bar\tau_\alpha:\Test{D}\rta\Lam{D}.$$

The first operation, \newsym{\protect\tau_\alpha}, will verify that its argument $M\in \Lam{D}$ has the point $\alpha$ in its interpretation. Intuitively, this is performed by recursively unfolding the B\"ohm tree of $M$ and succeeding ({\em i.e.}, converging) when $\alpha$ is in the interpretation of the finite unfolded B\"ohm tree. If $\alpha\not\in \llb M\rrb$, the test~$\tau_\alpha(M)$ will either diverge or refute (raising a $\0$ considered as an error). Concretely, it is an infinite application that feeds its argument with empty $\bar\tau$ operators.

The second operator, \newsym{\protect\bar\tau_\alpha}, simply constructs a term of interpretation $\da\alpha$ if its argument succeeds and diverges otherwise. Concretely, it is an infinite abstraction that runs its test argument, but also tests each of its applicants using $\tau$ operators.

In addition to these operators, we use \newdefprem{sums}{of terms} \newdefinvinv{sums}{of tests} and \newdefprem{products}{of tests} as ways to introduce may (for the addition) \newdefinvinv{may non-determinism}{in \Lam{D}} and must (for the multiplication) non-determinism\newdefinvinv{must non-determinism}{in \Lam{D}}; in the spirit of the $\lambda\+||$-calculus \cite{DLP98}. Indeed, these two forms of non-determinism are necessary to explore the branching of B\"ohm trees.

The idea of these two operators is to use the parametricity of our terms toward their intersection types. As a result, $\bar\tau_\alpha(\epsilon)$ (further on denoted $\bareps_\alpha$), that transfers the always succeeding test $\epsilon$ into a term of interpretation $\da\alpha$, constitutes the canonical term of type $\alpha$; its behavior is exactly the common behavior of every term of type $\alpha$. Symmetrically, the test $\tau_\alpha(M)$ will verify whether $M$ behaves like a term of type $\alpha$.

Hereafter, $D$ denotes a fixed extensional  \Kweb.

\begin{figure*} \caption{Grammar of the calculus with $D$-tests \label{fig:gr1}}
  \begin{center}
    \begin{tabular}{l l r @{\ ::=\ } l  l}
      (term) & 
      $\Lam{D}$ & 
      $M,N$ \hspace{1em} & 
      \hspace{1em} $x\quad |\quad \lambda x.M\quad |\quad M\ N\quad |\quad \newsym{\protect\sum_{i\le n}\protect\bar\tau_{\alpha_i}(Q_i)}$ & 
      \hspace{1em} $,\forall (\alpha_i)_i\in D^n,n\ge 0$ 
      \vspace{0.5em}\\
      (test) & 
      $\Test{D}$ & 
      $P,Q$ \hspace{1em} & 
      \hspace{1em} $\newsym{\protect\sum_{i\le n}P_i}\quad |\quad \newsym{\protect\prod_{i\le n}P_i}\quad |\quad \protect\tau_{\alpha}(M)$ & 
      \hspace{1em} $,\forall \alpha\in D,n\ge 0$ \\
    \end{tabular}
  \end{center}
\end{figure*}
\begin{definition}\label{def:LtauD-calculus}
  The \newdefpremsec{\Lcalcul}{with D-tests}, for short \newsym{\protect\Lam{D}}, is given by the grammar in Figure~\ref{fig:gr1}.
  We denote the empty sum by \newsym{\protect\0}, and the empty product by \newsym{\protect\eps}. Binary sums (resp. products) can be written with infix notation, {\em e.g.} $P\+Q$ \newsyminvinv{+}{between tests}\newsyminvinv{+}{between terms} (resp $P\pt Q$\newsyminvinv{\cdot}{between tests}).
  
  Moreover, we use the notation $\newsym{\protect\bareps_\alpha}:=\bar\tau_\alpha(\eps)$ and $\newsym{\protect\bareps_a}:=\sum_{\alpha\in a}\bareps_\alpha$; which are terms.
  
  Sums and products are considered as multisets, in particular we suppose associativity, commutativity and neutrality with, respectively, $\0$ and $\eps$. 

  In the following, an \newdef{abstraction} can refer either to a $\lambda$-abstraction or to a sum of $\bar\tau$ operators. This notation is justified by the behavior of $\Sigma_i\bar\tau_{\alpha_i}(Q_i)$ that mimics an infinite abstraction.
  
  The operational semantics is given by three sets of rules in Figure~\ref{fig:OS}. The {\em main rules} of Figure~\ref{fig:ER} are the effective rewriting rules. The {\em distributive rules} of Figure~\ref{fig:DS} implement the distribution of the sum over the test-operators and the product. The small step semantics \newdefinvinv{reduction}{for \Lam{D}} \newsymprem{\protect\rta}{as reduction in \Lam{D}} is the free contextual closure ({\em i.e.}, by the rules of Figure~\ref{fig:FCR}) of the rules of Figures~\ref{fig:ER} and ~\ref{fig:DS}. The {\em contextual rules} of Figure~\ref{fig:CR} implement the \newdefsecpreminv{reduction}{head}{for \Lam{D}} \newsymprem{\protect\rta_h}{for \Lam{D}} that is the specific contextual extension we are considering.
\end{definition}

  \begin{figure*} \caption{Operational semantics of the calculus with $D$-tests}\label{fig:OS}
    \vspace{0.5em}
    \begin{subfigure}[b]{1\textwidth}
      \hspace{-3em}
      \begin{center}
      \begin{tabular}{c l r c l}
        \vspace{1em}
        $(\beta)$\hspace{2em} &  & $(\lambda x.M)\ N$ & $\!\rta\!$ & $M[N/x]$\\
        \vspace{1em}
        \newsymsec{reduction rule}{(\protect\bar\tau)}\hspace{2em} & $\forall \beta_i = a_i\cons\alpha_i,$ & $(\Sigma_i\bar\tau_{\beta_i}(Q_i))\ N$ & $\!\rta\!$ & $\Sigma_i\bar\tau_{\alpha_i}(Q_i\ \pt\ \Pi_{\gamma\in a_i}\tau_\gamma(N))$\\
        \vspace{1em}
        \newsymsec{reduction rule}{(\protect\tau)}\hspace{2em} & $\forall \beta = a\cons\alpha,$\hspace{4em} & $\tau_\beta(\lambda x.M)$ & $\!\rta\!$ & $\tau_{\alpha}(M[\bareps_a/x])$\\
        \vspace{1em}
        \newsymsec{reduction rule}{(\protect\tau\protect\bar\tau)}\hspace{2em} & $\forall \alpha,\forall (\beta_i)_{i},$ & $\tau_\alpha(\Sigma_{i}\bar\tau_{\beta_i}(Q_i))$ & $\!\rta\!$ & $\Sigma_{\{i\mid \alpha\le \beta_i\}}Q_i$
      \end{tabular}
      \end{center}
      \vspace{-0.5em}
      \caption{Main rules}
      \label{fig:ER}
    \end{subfigure}

    \vspace{1em}
    \begin{subfigure}[b]{1\textwidth}
      \vspace{3em}
      \begin{center}
      \begin{tabular}{c r c l}
        \newsymsec{reduction rule}{(\protect\pt\+)}\hspace{4em} & $\Pi_{i\le n}\Sigma_{j\le k_i}Q_{i,j}$ &$\hspace{-0.5em}\rta\hspace{-0.5em}$ &$\Sigma_{j_1\le k_1,...,j_n\le k_n}\Pi_{i\le n}Q_{i,j_i}$
        \vspace{1em}\\
        \newsymsec{reduction rule}{(\protect\bar\tau+)}\hspace{4em} & $\bar\tau_\alpha(\Sigma_iQ_i)$ &$\hspace{-0.5em}\rta\hspace{-0.5em}$ &$\Sigma_i\bar\tau_\alpha(Q_i)$\\
      \end{tabular}
      \end{center}
      \caption{Distribution of the sum}
      \label{fig:DS}
    \end{subfigure}
   
    \vspace{1em}
    \begin{subfigure}[b]{1\textwidth}
      \vspace{3em}
      \begin{center}
        \AxiomC{$M\rta_h M'$}
        \RightLabel{\newsymsec{reduction rule}{(h\dash c\lambda)}}
        \UnaryInfC{$\lambda x.M\rta_h \lambda x.M'$}
        \DisplayProof\hskip 50pt
        \AxiomC{$M\rta_h M'$}
        \AxiomC{$M$ is an application}
        \RightLabel{\newsymsec{reduction rule}{(h\dash c\protect\at)}}
        \BinaryInfC{$M\ N\rta_h M'\ N$}
        \DisplayProof\\
        \vspace{1.3em}
        \AxiomC{$M\rta_h M'$}
        \AxiomC{$M$ is an application}
        \RightLabel{\newsymsec{reduction rule}{(h\dash c\protect\tau)}}
        \BinaryInfC{$\tau_\alpha(M)\rta_h \tau_\alpha(M')$}
        \DisplayProof\hskip 40pt
        \AxiomC{$Q\rta_h Q'$}
        \AxiomC{\hspace{-1em}$Q$ is not a sum}
        \RightLabel{\newsymsec{reduction rule}{(h\dash c\protect\bar\tau)}}
        \BinaryInfC{$\bar\tau_\alpha(Q)\rta_h \bar\tau_\alpha(Q')$}
        \DisplayProof\\
        \vspace{1.3em}
        \AxiomC{$M\rta_h M'$}
        \RightLabel{\newsymsec{reduction rule}{(h\dash cs)}}
        \UnaryInfC{$M+N\rta_h M'+N$}
        \DisplayProof\hskip 20pt
        \AxiomC{$Q\rta_h Q'$}
        \RightLabel{\newsymsec{reduction rule}{(h\dash c\protect\!+\protect\!)}}
        \UnaryInfC{$Q+P\rta_h Q'+P$}
        \DisplayProof\hskip 20pt
        \AxiomC{$Q\rta_h Q'$}
        \AxiomC{\hspace{-1em}$Q$ is not a sum}
        \RightLabel{\newsymsec{reduction rule}{(h\dash c\protect\pt)}}
        \BinaryInfC{$Q\pt P\rta_h Q'\pt P$}
        \DisplayProof
      \end{center}
      \caption{Contextual rules for the head reduction}
      \label{fig:CR}
    \end{subfigure}
    \begin{subfigure}[b]{1\textwidth}
      \vspace{4em}
      \begin{center}
        \AxiomC{$M\rta M'$}
        \RightLabel{\newsymsec{reduction rule}{(c\lambda)}}
        \UnaryInfC{$\lambda x.M\rta \lambda x.M'$}
        \DisplayProof\hskip 30pt
        \AxiomC{$M\rta M'$}
        \RightLabel{\newsymsec{reduction rule}{(c\protect\at L)}}
        \UnaryInfC{$M\ N\rta M'\ N$}
        \DisplayProof\hskip 30pt
        \AxiomC{$N\rta N'$}
        \RightLabel{\newsymsec{reduction rule}{(c\protect\at R)}}
        \UnaryInfC{$M\ N\rta M\ N'$}
        \DisplayProof\\
        \vspace{1.3em}
        \AxiomC{$M\rta M'$}
        \RightLabel{\newsymsec{reduction rule}{(c\protect\tau)}}
        \UnaryInfC{$\tau_\alpha(M)\rta \tau_\alpha(M')$}
        \DisplayProof\hskip 40pt
        \AxiomC{$Q\rta Q'$}
        \RightLabel{\newsymsec{reduction rule}{(c\protect\bar\tau)}}
        \UnaryInfC{$\bar\tau_\alpha(Q)\rta \bar\tau_\alpha(Q')$}
        \DisplayProof\\
        \vspace{1.3em}
        \AxiomC{$M\rta M'$}
        \RightLabel{\newsymsec{reduction rule}{(cs)}}
        \UnaryInfC{$M+N\rta M'+N$}
        \DisplayProof\hskip 20pt
        \AxiomC{$Q\rta Q'$}
        \RightLabel{\newsymsec{reduction rule}{(c\protect\!+\protect\!)}}
        \UnaryInfC{$Q+P\rta Q'+P$}
        \DisplayProof\hskip 20pt
        \AxiomC{$Q\rta Q'$}
        \RightLabel{\newsymsec{reduction rule}{(c\protect\pt)}}
        \UnaryInfC{$Q\pt P\rta Q'\pt P$}
        \DisplayProof
      \end{center}
      \caption{Contextual rules for the full reduction}
      \label{fig:FCR}
    \end{subfigure}
  \end{figure*}


\begin{example}
  The operational behavior of $D$-tests depends on $D$. Recall the \Kwebs of Example~\ref{example:1}. In the case of Scott's $D_\infty$ we have in $\Lamb_{\tau(D_\infty)}$:
  \begin{align*}
    \tau_*(\underline{(\lambda xy.x\ y)\ \bareps_*}) \quad 
      & \stackrel{\beta}{\rta}_h\quad \underline{\tau_*(\lambda y}.\bareps_*\ y) \quad 
        \stackrel{\tau}{\rta}_h \quad \tau_*(\underline{\bareps_*\ \bareps_{\emptyset}})  \\
      & \stackrel{\bar\tau}{\rta}_h \quad \tau_*(\underline{\bareps_*}) \quad
        = \quad \underline{\tau_*(\bar\tau_*}(\eps)) \quad
        \stackrel{\tau\bar\tau}{\rta}_h \quad \eps, \\
        \\
    \tau_*(\underline{(\lambda xy.y\ x)\ \bareps_*}) \quad
      & \stackrel{\beta}{\rta}_h \quad \underline{\tau_*(\lambda y}.y\ \bareps_*) \quad
        \stackrel{\tau}{\rta}_h \quad \tau_*(\underline{\bareps_{\emptyset}}\ \bareps_*)\\ 
      & = \quad \tau_*(\underline{\0\ \bareps_*}) \quad
        \stackrel{\bar\tau}{\rta}_h \quad \underline{\tau_*(\0)} \quad
        \stackrel{\tau\bar\tau}{\rta}_h \quad \0.
  \end{align*}
  In the case of Park $P_\infty$:
  \begin{align*}
    \underline{\tau_*(\lambda x}.xx) \quad  
      & \stackrel{\tau}{\rta}_h \quad \tau_*(\underline{\bareps_*\ \bareps_*}) \quad
        \stackrel{\bar\tau}{\rta}_h \quad \underline{\tau_*(\bar\tau_*}(\underline{\tau_*(\bareps_*}))) \quad
        \stackrel{\tau\bar\tau}{\rta}_h\stackrel{\tau\bar\tau}{\rta}_h \quad \eps.
  \end{align*}
  In the case of Norm:
  \begin{align*}
    \underline{\tau_p(\lambda x}.x) \quad
      & \stackrel{\tau}{\rta}_h \quad \underline{\tau_p(\bareps_q)} \quad 
        \stackrel{\tau\bar\tau}{\rta}_h \quad \eps, 
    &\underline{\tau_q(\lambda x}.x) \quad
      & \stackrel{\tau}{\rta}_h \quad \underline{\tau_q(\bareps_p)} \quad
        \stackrel{\tau\bar\tau}{\rta}_h \quad \0.
  \end{align*}
\end{example}

\begin{example}\label{ex:barepsM1...Mn}
  In any K-model $D$, given $\alpha=a_1\cons\cdots\cons a_{n+1}\cons\beta\in D$, and if we denote \linebreak $\alpha'=a_2\cons\cdots\cons a_{n+1}\cons\beta$ we have:
  \begin{eqnarray*} \ 
    \bareps_\alpha\ M_1 \cdots M_{n+1} & \ruledarrow{\bar\tau}{h}{} & \bar\tau_{\alpha'}(\Pi_{\gamma\in a_1}\tau_\gamma(M_1))\ M_2 \cdots M_{n+1} \\
    & \ruledarrow{\bar\tau}{h}{n} & \bar\tau_{\beta}(\Pi_{i\le n+1}\Pi_{\gamma\in a_i}\tau_\gamma(M_i))
  \end{eqnarray*}
\end{example}

\begin{remark}\label{rk:polarisedTests}
  In a polarized (or classical) framework with explicit co-terms (or stacks) as the framework presented in \cite{Mun09}, tests would correspond to commands (or processes), or, more exactly, to conjunctions and disjunctions of commands. Indeed, a test $\tau_\alpha(M)$ is nothing else than the command $\langle M\mid\pi_\alpha\rangle$ where $\pi_\alpha$ would be the canonical co-term of interpretation $\ua\alpha$, the same way that $\bareps_\alpha$ is the canonical term of interpretation $\da\alpha$. Similarly, the term $\bar\tau(Q)$ can be seen as the canonical term $\bareps_\alpha$ endowed with a parallel composition referring to the set of commands $Q$. To resume, we have:
  \begin{align*}
    \tau_\alpha(M)\ &\simeq\  \langle M \mid \ua\alpha \rangle   &   \langle \bar\tau_\alpha(Q)\mid\pi\rangle\ &\simeq\ \langle \da\alpha \mid \pi\rangle\pt Q
  \end{align*}
\end{remark}

\begin{remark}
  In the conference version \cite{Bre14}, the rule $(\tau\bar\tau)$ is decomposed into three rules (the distribution of the sum over $\tau$, denoted $(\tau+)$ and two versions of $(\tau\bar\tau)$ depending on whether $\alpha\le \beta$). This decomposition was easier to understand as more atomic, but ultimately it always reproduces our actual rule $(\tau\bar\tau)$ and does not permit to use Theorem~\ref{th:invHeadReducibility}.
\end{remark}

\begin{prop}\label{prop:TestHNF}
A test is in {\em head-normal form} iff it has the following shape:
$$\Sigma_{i\le k}\Pi_j\tau_{\alpha_{i,j}}(x_{i,j}\ M^1_{i,j}\cdots\ M^n_{i,j}),$$
with $k\ge 1$ and $M^k_{i,j}$ any term.\\
A term is in \newdefprem{head-normal form}{for \Lam{D}} if it has one of the following shapes:
\begin{align*}
\lambda x_1....x_n.&y\ M_1 \cdots\ M_m,    &
\text{or}\hspace{1.5em} \lambda x_1...x_n.& \Sigma_{i\le k}\bar\tau_{\alpha_i}(Q_i),
\end{align*}
where $m,n\ge 0$, $k\ge 1$, $(\alpha_i)_i\in D^k$,  $M_i$ is any term, and every $Q_i$ any test in head-normal form without sums.
\end{prop}
\begin{proof}
By structural induction on the grammar of $\Lam{D}$. In particular, notice that any test of the shape $\tau_\alpha(\lambda x.M)$ is not a head-normal form because $i_D$ is surjective and thus $\alpha=a\cons\beta$ for some $a, \beta$ and we can apply Rule $(\tau)$.
\end{proof}


\begin{definition}
  A term (resp. test) is {\em head-converging} \newdefinvinv{head convergence}{in \Lam{D}} if it head reduces to a \newdef{may-head-normal form} (denoted \newsym{\mathrm{mhnf}}) that is either a head-normal form or a term (resp. test) of the form
  \begin{align*}
    \lambda x_1...x_n. (\bar\tau_{\alpha}(Q)+N)&  &
    \text{resp. }& Q_1+Q_2
  \end{align*}
with $\bar\tau_\alpha(Q)$ (resp. $Q_1$) in head-normal form and $N$ any term (resp. $Q_2$ any test). This corresponds to a {\em may-convergence} for the sum. Coherently with the head convergence in \Lcalcul, the convergence will be denoted by \newsymprem{\protect\Da}{in \Lam{D}}\newsyminvinv{\protect\Da N}{in \Lam{D}} and the divergence by \newsymprem{\protect\Ua}{in \Lam{D}}.
\end{definition}

\begin{example}
  For any $n\in\Nat$, the term $\underline{n}\ (\lambda x. \bar\tau_\alpha(\tau_\alpha(x)\+\tau_\beta(x)))\ \bareps_\alpha$ may-head-converges.
\end{example}

\todo{see comments}

Let us notice that this calculus enjoys the properties of confluence and standardization (Th.~\ref{th:confluence} and Th.~\ref{th:standardisation}). We also have another syntactical theorem stating invariance wrt the head-convergence in at most $n$ steps, denoted  $\Da_{\!\!n}$ (Theorem~\ref{th:invHeadReducibility}). This means that performing a non-head reduction can only reduce the length of convergence.

%
%

\begin{definition}
Grammars of \newdef{term-context}s \Lamcont{D} and \newdef{test-context}s \Testc{D} are given in Figure~\ref{fig:grC}. 
\end{definition}

\begin{figure*}
  \caption{Grammar of the contexts in a calculus with $D$-tests \label{fig:grC}}
  \begin{tabular}{l l r @{\ ::=\ } l l}
    \!\!(term-context)\!\! &
    \newsym{\protect\Lamcont{D}} & 
    $C$  & 
    $x\quad |\quad \newsymprem{\protect\llc.\protect\rrc}{for \protect\Lam{D}}\quad |\quad C \ C' \quad |\quad \lambda x.C \quad |\quad \sum_{i\le n}\bar\tau_{\alpha_i}(K_i )$ & 
    $,\forall (\alpha_i)_i\in D^n,n\ge 0$ 
    \vspace{0.5em}\\
    \!\!(test-context)\!\!  & 
    \newsym{\protect\Testc{D}}  & 
    $K$  & 
    $\sum_{i\le n}K_i \quad |\quad \prod_{i\le n}K_i \quad |\quad \tau_{\alpha}(C )$ & 
    $,\forall \alpha\in D,n\ge 0$ \\
  \end{tabular}
\end{figure*}

\begin{definition}
The \newdefprem{observational preorder}{of \Lam{D}} \newsym{\protect\leobtau{D}} of \Lam{D} is defined by: \vspace{-0.5em}
 $$M\sqsubseteq_{\tau(D)}N\ \ \text{iff}\ \ (\forall K \mathrm{\in} \Testc{D},\ K\llc M\rrc\Da\ \text{ implies }\ K\llc N\rrc\Da).$$
We denote by \newsym{\protect\equivobtau{D}} the \newdefprem{observational equivalence}{of \Lam{D}}, {\em i.e.}, the equivalence induced by \leobtau{D}.
\end{definition}

\begin{remark}
  The observational preorder could have been defined using term-contexts rather than test-contexts, but this appears to be equivalent and test-contexts are easier to manipulate (because normal forms for tests are simpler).
  \begin{proof}
    For any test $Q$ and for any $\alpha$, $Q\Da$ iff $\bar\tau_\alpha(Q)\Da$. Conversely, for all $M$, there is $n\in \Nat$ and~$\alpha\in D$ such $M\Da$ iff $\tau_{\alpha}(M x_0 \underset{n}{\cdots} x_0)\Da$ (remark that if $N$ diverges, then $\tau_{\alpha}(N \underbrace{x_0 {\cdots} x_0}_{n\text{ times}})\Ua$).
  \end{proof}
\end{remark}
\todo{can be removed}

\subsection{Semantics}\label{subsubsec:TestSemantics}\ \newline
The standard interpretation of $\Lamb$ into $D$ (Fig.~\ref{fig:intLam} and recalled here in Figure~\ref{fig:interpret}) can be extended to~$\Lam{D}$ (Fig.~\ref{fig:intTests}). \newdefinvinv{interpretation}{of \Lam{D} in $D$} \newdefinvinvinv{interpretation}{of \Lamb}{in a \Kweb} \newsyminvinv{\protect\llb.\protect\rrb^{\protect\vec x}}{for \Lam{D} in $D$}
\begin{figure*} \label{fig:TestInter}
  \vspace{-0.5em}
  \caption{Direct interpretation in $D$ \label{fig:interpret}}
  \vspace{1em}
  \hspace{-1em}
  \begin{subfigure}[b]{1\textwidth}
    \begin{center}
      $\llb x_i \rrb_D^{\vec x}  =  \{(\vec a,\alpha)\ |\ \alpha\le\beta\in a_i\}$
      \hspace{4em}
      $ \llb \lambda y.M \rrb_D^{\vec x}  =  \{(\vec a,(b\cons\alpha))\ |\ (\vec ab,\alpha)\in\llb M\rrb_D^{\vec xy}\}$
      \vspace{0.3em}\\
      $\!\llb M\ N \rrb_D^{\vec x}  =  \{(\vec a,\alpha)\ |\ \exists b,(\vec a,(b\cons\alpha))\in\llb M\rrb_D^{\vec x}\ \wedge \forall \beta\mathrm{\in} b, (\vec a,\beta)\in\llb N\rrb_D^{\vec x}\}$
    \end{center}
    \caption{Interpretation of $\Lamb$}
  \end{subfigure}

  \vspace{0.5em}
  \begin{subfigure}[b]{1\textwidth}
    \begin{center}
      $\llb \Sigma_{i\le k}\bar\tau_{\alpha_i}(Q_i) \rrb_D^{\vec x}  =  \bigcup_{i\le k}\{(\vec a,\beta)\ |\ \vec a\in \llb Q_i\rrb_D^{\vec x}\ \wedge\ \beta\le_{D}\alpha_i\}$
      \hspace{1em}
      $\llb \0 \rrb^{\vec x}_D  =  \emptyset$
      \vspace{0.3em}\\
      $\llb \tau_{\alpha}(M) \rrb_D^{\vec x}  =  \{\vec a\ |\ (\vec a,\alpha)\in\llb M\rrb_D^{\vec x}\}$
      \vspace{0.3em}\\
      $\llb \Pi_{i\le k}Q_i \rrb_D^{\vec x}  =  \bigcap_{i\le k} \llb Q_i\rrb_D^{\vec x}$
      \hspace{1em}
      $\llb \eps \rrb^{\vec x}_D  =  \Achf{D}^{\vec x}$
      \hspace{1em}
      $\llb \Sigma_{i\le k}Q_i \rrb_D^{\vec x}  =  \bigcup_{i\le k}\llb Q_i\rrb_D^{\vec x}$
      \hspace{1em}
      $\llb \0 \rrb^{\vec x}_D  =  \emptyset$
    \end{center}
    \caption{Interpretation of tests extensions \label{fig:intTests}}
  \end{subfigure}
\end{figure*}

\begin{definition}
  A term $M$ with $n$ free variables is {\em interpreted} as a morphism (Scott-continuous function) from $D^n$ to $D$ and a test $Q$ with $n$ free variables as a morphism from $D^n$ to the dualizing object $\{*\}$ (singleton poset):
  \begin{align*}
    \llb M\rrb_D^{x_1,...,x_n}\ &\subseteq\ (D\Rta\!\cdots\Rta D\Rta D)\ \simeq\ (\Achf{D}^{op})^n\times D\\
    \llb Q\rrb_D^{x_1,...,x_n}\ &\subseteq\ (D\Rta\!\cdots\Rta D\Rta\{*\})\ \ \simeq\ (\Achf{D}^{op})^n
  \end{align*}
  This interpretation is given in Figure~\ref{fig:interpret} by structural induction.
\end{definition}

\begin{prop}
  For any extensional \Kweb $D$, $D$ is a model of the \Lcalcul with $D$-tests, {\em i.e.}, the interpretation is invariant under reduction.
\end{prop}
\begin{proof}
The invariance under $\beta$-reduction  is obtained, as usual, by the Cartesian closedness of \Scottlb. The other rules are easy to check directly.
\end{proof}

\begin{prop} \label{prop:contInvar}
  For any extensional \Kweb $D$, the interpretation is invariant by context, \linebreak {\em i.e.},~$\llb M\rrb^{\vec x}=\llb N\rrb^{\vec x}$ implies that for any test/term-context $C$, $\llb C\llc M\rrc\rrb^{\vec x}=\llb C\llc N\rrc\rrb^{\vec x}$.
\end{prop}
\begin{proof}
  By easy induction on $C$.
\end{proof}

The idea of intersection types can be generalized to $\Lam{D}$. We introduce in Figure~\ref{fig:tyTests} \newsyminvinv{\Gamma\vdash M:\tau}{intersection type from a \protect\Kweb} a type assignment system associating with any term $M\in \Lam{D}$ an element of $D$ under an environment~$(x_i{:}a_i)_i$ with $a_i\in \Achf{D}$. The following theorem gives the equivalence between the interpretation of a term and the set of judgments derivable from the type system. 

\begin{theorem}[Intersection types]
  Let $M$ be a term of $\Lam{D}$,  (resp. $Q$ be a test of $\Test{D}$), the following statements are equivalent:
  \begin{itemize}
  \item $(\vec a,\alpha)\in\llb M\rrb_D^{\vec x}$ (resp. $\vec a\in \llb Q\rrb_D^{\vec x}$),
  \item the type judgment $\vec x:\vec a\vdash M:\alpha$ (resp. $\vec x:\vec a\vdash Q$) is derivable by the rules of Figure~\ref{fig:tyTests}.
  \end{itemize}
\end{theorem}
\begin{proof}
By structural induction on the grammar of $\Lam{D}$.
\end{proof}

\begin{figure*}
  \begin{center}
    \AxiomC{$\alpha\in a$}
    \UnaryInfC{$x:a\vdash x:\alpha$}
    \DisplayProof\hskip 60pt
    \AxiomC{$\Gamma\vdash M:\alpha$}
    \UnaryInfC{$\Gamma,x:a\vdash M:\alpha$}
    \DisplayProof\hskip 60pt
    \AxiomC{$\Gamma\vdash M:\beta$}
    \AxiomC{$\alpha\le\beta$}
    \BinaryInfC{$\Gamma\vdash M:\alpha$}
    \DisplayProof\\ \vspace{1em}
    \AxiomC{$\Gamma,x:a\vdash M:\alpha$}
    \UnaryInfC{$\Gamma\vdash \lambda x.M:a\cons\alpha$}
    \DisplayProof\hskip 70pt
    \AxiomC{$\Gamma\vdash M:a\cons\alpha$}
    \AxiomC{$\forall \beta\in a,\ \Gamma\vdash N:\beta$}
    \BinaryInfC{$\Gamma\vdash M\ N:\alpha$}
    \DisplayProof\\ \vspace{1em}
    \AxiomC{$\exists i\le n,\ \Gamma\vdash Q_i$}
    \UnaryInfC{$\Gamma\vdash \Sigma_{i\le n}\bar\tau_{\alpha_i}(Q_i):\alpha_i$}
    \DisplayProof\hskip 25pt
    \AxiomC{$\Gamma\vdash M:\alpha$}
    \UnaryInfC{$\Gamma\vdash \tau_\alpha(M)$}
    \DisplayProof\hskip 25pt
    \AxiomC{$\exists i\le n,\ \Gamma\vdash Q_i$}
    \UnaryInfC{$\Gamma\vdash \Sigma_{i\le n}Q_i$}
    \DisplayProof\hskip 25pt
    \AxiomC{$\forall i\le n,\ \Gamma\vdash Q_i$}
    \UnaryInfC{$\Gamma\vdash \Pi_{i\le n}Q_i$}
    \DisplayProof
  \end{center}
  \caption{Intersection type system associated with tests extensions \label{fig:tyTests}}
\end{figure*}

\begin{remark}\label{rk:typ_sub}
  In particular, an easy induction gives that if $\vdash M[N/x]:\alpha$ then there is $a$ such that~$N:a\vdash M:\alpha$.
\end{remark}

\todo{ref to a forthcoming article?}

\subsubsection{Full abstraction and sensibility for tests}\label{subsubsec:FAwT}\ \newline
The main theorem (Th.~\ref{th:final}) uses the assumption of sensibility of $D$ for \Lam{D}. The sensibility is simply asking for the diverging terms $M\in\Lam{D}$ to have empty interpretation as specified in Definition~\ref{def:sensibilityWTests}. Its interest is in implying directly the inequational full abstraction of $D$ for \Lam{D} ({\em i.e.} for its observational preorder) as we will see in Theorem~\ref{th:FAwT}. The proof of Theorem~\ref{th:FAwT} needs a technical counterpart that is basically the \newdef{definability} of \Lam{D} stated in Theorem~\ref{th:caracFAwT}. This definability theorem is not usual and appears to be stronger and more useful for future developments.

First we recall the definition of sensibility:

\begin{definition}\label{def:sensibilityWTests}
  An extensional \Kweb $D$ is {\em sensible for \Lam{D}} \newdefinvinv{sensibility}{of $D$ for \Lam{D}} whenever diverging terms (resp. tests) correspond exactly to the terms (resp. tests) having empty interpretation, {\em i.e.}, for all $M\in\Lam{D}$ and $Q\in \Test{D}$:
  \begin{align*}
    M\Ua &\quad \Lra \quad \llb M\rrb_D^{\vec x}=\emptyset
    & Q\Ua &\quad \Lra \quad \llb Q\rrb_D^{\vec x}=\emptyset
  \end{align*}
\end{definition}

\begin{lemma}\label{lemma:test}
  If $D$ is sensible for $\Lamb_{\tau(D)}$ then:
  $$(\vec a b,\alpha)\in\llb M\rrb^{\vec y x}\ \ \Lra\ \ \ (\vec a,\alpha)\in\llb M[\bareps_b/x]\rrb^{\vec y},$$
  \vspace{-1em}
  $$(\vec a, \alpha)\in\llb M\rrb^{\vec y}\ \ \Lra\ \ \vec a\in\llb \tau_\alpha(M)\rrb^{\vec y}.$$
\end{lemma}
\begin{proof}
 This lemma and its test counterpart is proved by a straightforward induction on $M$ (and $Q$ of the test version).
\end{proof}

\begin{theorem}[Definability]\label{th:caracFAwT}
  If $D$ is sensible for $\Lamb_{\tau(D)}$ then:
  $$ (\vec a,\alpha)\in\llb M\rrb^{\vec x}\ \ \Lra\ \ \tau_\alpha(M[(\bareps_{a_i}/x_i)_{i\le n}]) \Da.$$
\end{theorem}
\begin{proof}
  If $(\vec a,\alpha)\in\llb M\rrb^{\vec x}$ then $\llb\tau_\alpha(M[(\bareps_{a_i}/x_i)_{i\le n}])\rrb$ is not empty by Lemma~\ref{lemma:test}, thus it converges by sensibility. Conversely, if $\tau_\alpha(M[(\bareps_{a_i}/x_i)_{i\le n}])\Da$ then its interpretation is non empty, which means that in particular~$*{\in}\llb\tau_\alpha(M[(\bareps_{a_i}/x_i)_{i\le n}])\rrb$ (where $*$ denotes the only inhabitant of $\bot$) and thus, by Lemma~\ref{lemma:test}, $(\vec a,\alpha)\in\llb M\rrb^{\vec x}$.
\end{proof}

\begin{theorem}[full abstraction]\label{th:FAwT}
  For any extensional \Kweb $D$, if $D$ is sensible for $\Lam{D}$, then $D$ is inequationally fully abstract for the observational preorder of \Lam D: \newdefinvinv{inequational full abstraction}{of $D$ for \Lam{D}}
  $$ \llb M\rrb\subseteq \llb N\rrb\quad \Lra\quad \forall C \in \Testc{D}, C\llc M\rrc\Da\Rta C\llc N\rrc\Da.$$
\end{theorem}
\begin{proof}
  Let $\llb M\rrb\subseteq \llb N\rrb$ and $C\llc M\rrc\Da$. Then by sensibility we have that $\llb C\llc M\rrc\rrb$ is non-empty. Moreover, by Proposition~\ref{prop:contInvar} we have that $\llb C\llc M\rrc\rrb\subseteq \llb C\llc N\rrc\rrb$. Thus $\llb C\llc N\rrc\rrb$ is non-empty and by sensibility, $C\llc N\rrc\Da$.\todo{change in the thesis!!!}\\
  Conversely, suppose that for all context $C \in \Testc{D}, C\llc M\rrc\Da\Rta C\llc N\rrc\Da$ and let $(\vec a,\alpha)\in\llb M\rrb^{\vec x}$:\\
  Then by Theorem~\ref{th:caracFAwT}, $\tau_\alpha(M[(\bareps_{a_i}/x_i)_{i\le n}]) \Da$ where $n$ is the length of $\vec a$. Thus, after stating the context $C=\tau_\alpha((\lambda x_...x_n.\llc.\rrc)\ \bareps_{a_1}\cdots\bareps_{a_n})$, we have  $C\llc M\rrc\rta_h^n\tau_\alpha(M[(\bareps_{a_i}/x_i)_{i\le n}])\Da$ which implies that $C\llc N\rrc\Da$. However, there is no choice\footnote{We have to verify that this are the only possible reductions because in general the head-reduction is not determonistic in $\Lam D$.} for the $n$ first head reductions of $C\llc N\rrc$, those are forced to be $C\llc N\rrc\rta^n_h\tau_\alpha(N[(\bareps_{a_i}/x_i)_{i\le n}])$ so that this term is head-converging. Then by applying the reverse implication of Theorem~\ref{th:caracFAwT} we conclude $(\vec a,\alpha)\in \llb N\rrb^{\vec x}$.
\end{proof}

  \subsection{Technical theorems}
%
\subsubsection{Confluence}\ \newline
This section is dedicated to the proof of Theorem~\ref{th:confluence} stating the confluence of the reduction $\rta$ in \Lam{D}. The proof uses the diamond property of the full parallel reduction, following the proof of~\cite{Tait67} for the \Lcalcul.

 We define first the \newdefsecpreminv{reduction}{parallel}{with tests} \newsympreminvinv{\protect\Rta}{parallel reduction}{with tests}  in Figure~\ref{fig:PR}, allowing the parallel reduction of independent redexes.

  \begin{figure*}
    \vspace{1em}
    \begin{subfigure}[b]{1\textwidth}
      \hspace{-3em}
      \begin{center}
        \AxiomC{$M\Rta M'$}
        \AxiomC{$N\Rta N'$}
        \RightLabel{\newsymsec{reduction rule}{(P\protect\dash\beta)}}
        \BinaryInfC{$(\lambda x.M)\ N\Rta M'[N'/x]$}
        \DisplayProof\hskip 40pt
        \AxiomC{$M\Rta M'$}
        \AxiomC{$\forall i,\ Q_{i}\Rta \Sigma_j Q'_{ij}$}
        \RightLabel{\newsymsec{reduction rule}{(P\protect\dash \protect\bar\tau)}}
        \BinaryInfC{$\Sigma_i\bar\tau_{a_i\cons\alpha_i}(Q_{i})\ M\Rta\Sigma_{ij}\bar\tau_{\alpha_i}(Q'_{ij}\ \pt\ \Pi_{\gamma\in a_i}\tau_\gamma(M'))$}
        \DisplayProof\\
        \vspace{0.7em}
        \AxiomC{$M\Rta M'$}
        \RightLabel{\newsymsec{reduction rule}{(P\protect\dash \tau)}}
        \UnaryInfC{$\tau_{a\cons\alpha}(\lambda x.M)\Rta \tau_{\alpha}(M'[\boldsymbol{\bar\epsilon}_a/x])$}
        \DisplayProof\hskip 40pt
        \AxiomC{$\forall_i,Q_i\Rta Q'_i$}
        \RightLabel{\newsymsec{reduction rule}{(P\protect\dash \tau\protect\bar\tau)}}
        \UnaryInfC{$\tau_\alpha(\Sigma_i\bar\tau_{\beta_i}(Q_i))\Rta \Sigma_{\{i\mid \alpha\le \beta_i\}}Q'_i$}
        \DisplayProof
      \end{center}
    \caption{Main rules}
      \label{fig:PER}
    \vspace{1em}
    \end{subfigure}
    \vspace{1em}
    \begin{subfigure}[b]{1\textwidth}
      \begin{center}
        \AxiomC{$\forall i,\ Q_i\Rta Q_i'$}
        \RightLabel{\newsymsec{reduction rule}{(P\protect\dash \protect\bar\tau +)}}
        \UnaryInfC{$\bar\tau_{\alpha}(\Sigma_iQ_i)\Rta \Sigma_i\bar\tau_{\alpha}(Q'_i)$}
        \DisplayProof\hskip 40pt
        \AxiomC{$\forall ij,\ Q_{ij}\Rta Q'_{ij}$}
        \RightLabel{\newsymsec{reduction rule}{(P\protect\dash \protect\pt +)}}
        \UnaryInfC{$\Pi_{i\le n}\Sigma_{j\le k_i}Q_{ij}\Rta \Sigma_{j_1\le k_1,...,k_n\le k_n}\Pi_{i\le n}Q'_{ij_i}$}
        \DisplayProof
      \end{center}
      \caption{Distribution of the sum}
      \label{fig:PDS}
    \end{subfigure}
    \vspace{1em}
    \begin{subfigure}[b]{1\textwidth}
      \vspace{0.5em}
      \begin{center}
        \AxiomC{}
        \RightLabel{\newsymsec{reduction rule}{(P\protect\dash id)}}
        \UnaryInfC{$x\Rta x$}
        \DisplayProof\hskip 40pt
        \AxiomC{$M\Rta M'$}
        \RightLabel{\newsymsec{reduction rule}{(P\protect\dash c\lambda)}}
        \UnaryInfC{$\lambda x.M\Rta \lambda x.M'$}
        \DisplayProof\hskip 40pt
        \AxiomC{$M\Rta M'$}
        \AxiomC{$N\Rta N'$}
        \RightLabel{\newsymsec{reduction rule}{(P\protect\dash c\protect\at)}}
        \BinaryInfC{$M\ N\Rta M'\ N'$}
        \DisplayProof\\
        \vspace{0.5em}
        \AxiomC{$M\Rta M'$}
        \RightLabel{\newsymsec{reduction rule}{(P\protect\dash c\tau)}}
        \UnaryInfC{$\tau_\alpha(M)\Rta \tau_\alpha(M')$}
        \DisplayProof\hskip 60pt
        \AxiomC{$\forall i,\ M_i\Rta M'_i$}
        \RightLabel{\newsymsec{reduction rule}{(P\protect\dash cs)}}
        \UnaryInfC{$\Sigma_iM_i\Rta \Sigma_iM'_i$}
        \DisplayProof
      \end{center}
      \vspace{-0.5em}
      \caption{Contextual rules}
      \label{fig:PCR}
    \end{subfigure}
    \caption{Operational Semantics of parallel reduction}\label{fig:PR}
  \end{figure*}

\noindent\begin{lemma} \label{lemma:Rtarta}
  If $M\Rta N$ then $M\rta^* N$ and if $M\rta^* N$ then $M\Rta^* N$.\\
  In particular we have $\Rta^*=\rta^*$.
\end{lemma}
\begin{proof}
  Firstly remark that $\Rta$ is reflexive. Indeed, when we proceed by induction the only difficult case is $\epsilon\Rta \epsilon$ that is obtained by Rule $(P\dash\pt+)$ for $n=0$.\\
  Rules with similar names are then simulating each other except for
  \begin{itemize}
    \item $(c\at L)$ and $(c\at R)$ that are simulated by $(P\dash c\at)$.
    \item $(P\dash id)$ that is simulated by $\rta^\epsilon$ (the reduction in $0$ step).
    \item $(c+)$ that is a particular case of $(P\dash \pt+)$ with $n=1$ and $k_1=2$.
    \item $(c\pt)$ that is a particular case of $(P\dash \pt+)$ with $n=2$ and $k_1=k_2=1$.
    \item $(c\bar\tau)$ that is a particular case of $(P\dash \bar\tau+)$ where the sum has one element.
    \end{itemize}
  \end{proof}

For a term $M$ (resp. a test $Q$) we define the \newdef{maximal parallel reduct} \newsym{\protect M^+} (resp. \newsym{\protect Q^+}) by induction on $M$ and $Q$ in Figure~\ref{fig:def+}. Recall that by abstractions, we not only mean $\lambda$-abstractions, but also terms of the form $\Sigma_i\bar\tau_{\alpha_i}(Q_i)$.

\begin{figure*}
    \vspace{1em}
    \begin{subfigure}[b]{1\textwidth}
      \hspace{-3em}
      \begin{center}
        \AxiomC{$$}
        \RightLabel{\newsymsec{reduction rule}{(T\protect\dash \beta)}}
        \UnaryInfC{$((\lambda x.M)\ N)^+:= M^+[N^+/x]$}
        \DisplayProof\hskip 15pt
        \AxiomC{$\forall i,\quad Q_i^+=\Sigma_jQ'_{ij}$}
        \AxiomC{$\forall j,\ Q'_{i,j}$ are not sums\!\!}
        \RightLabel{\newsymsec{reduction rule}{(T\protect\dash \protect\bar\tau)}}
        \BinaryInfC{$((\Sigma_i\bar\tau_{a_i\cons\alpha_i}(Q_{i}))\ M)^+:=\Sigma_{ij}\bar\tau_{\alpha_i}(Q'_{ij}\ \pt\ \Pi_{\gamma\in a_i}\tau_\gamma(M^+))$}
        \DisplayProof\\
        \vspace{1em}
        \AxiomC{}
        \RightLabel{\newsymsec{reduction rule}{(T\protect\dash \tau)}}
        \UnaryInfC{$\tau_{a\cons\alpha}(\lambda x.M)^+:= \tau_{\alpha}(M^+[\boldsymbol{\bar\epsilon}_a/x])$}
        \DisplayProof\hskip 30pt
        \AxiomC{$\forall i\in I,\ \alpha\le_D\beta_i$}
        \AxiomC{$\forall i\in J,\ \alpha\not\le_D\beta_i$}
        \RightLabel{\newsymsec{reduction rule}{(T\protect\dash \tau\protect\bar\tau)}}
        \BinaryInfC{$\tau_\alpha(\Sigma_{i\in I\cup J}\bar\tau_{\beta_i}(Q_i))^+:= \Sigma_{i\in I}Q_i^+$}
        \DisplayProof
      \end{center}
    \caption{Main rules}
      \label{fig:fPER}
    \vspace{1em}
    \end{subfigure}
    \vspace{1em}
    \begin{subfigure}[b]{1\textwidth}
      \begin{center}
        \AxiomC{$\forall i,\ Q_i$ are not sums}
        \RightLabel{\newsymsec{reduction rule}{(T\protect\dash \protect\bar\tau+)}}
        \UnaryInfC{$\bar\tau_{\alpha}(\Sigma_iQ_i)^+:= \Sigma_i\bar\tau_{\alpha}(Q_i^+)$}
        \DisplayProof\hskip 40pt
        \AxiomC{$n\neq 1$ or $k_1\neq 1$}
        \AxiomC{the $Q_{ij}$ are not sums}
        \RightLabel{\newsymsec{reduction rule}{(T\protect\dash \protect\pt+)}}
        \BinaryInfC{$(\Pi_{i\le n}\Sigma_{j\le k_i}Q_{ij})^+:= \Sigma_{j_1\le k_1,...,k_n\le k_n}\Pi_{i\le n}Q_{ij_i}^+$}
        \DisplayProof
      \end{center}
      \caption{Distribution of the sum}
      \label{fig:fPDS}
    \end{subfigure}
    \vspace{1em}
    \begin{subfigure}[b]{1\textwidth}
      \vspace{0.5em}
      \begin{center}
        \AxiomC{}
        \RightLabel{\newsymsec{reduction rule}{(T\protect\dash id)}}
        \UnaryInfC{$x^+:=x$}
        \DisplayProof\hskip 20pt
        \AxiomC{}
        \RightLabel{\newsymsec{reduction rule}{(T\protect\dash c\lambda)}}
        \UnaryInfC{$(\lambda x.M)^+\Rta \lambda x.M^+$}
        \DisplayProof\hskip 20pt
        \AxiomC{$M$ is not an abstraction}
        \RightLabel{\newsymsec{reduction rule}{(T\protect\dash c\protect\at)}}
        \UnaryInfC{$(M\ N)^+:= M^+\ N^+$}
        \DisplayProof\\
        \vspace{0.7em}
        \AxiomC{$M$ is not an abstraction}
        \RightLabel{\newsymsec{reduction rule}{(T\protect\dash c\tau)}}
        \UnaryInfC{$\tau_\alpha(M):= \tau_\alpha(M^+)$}
        \DisplayProof\hskip 20pt
        \AxiomC{$k\neq 1$}
        \RightLabel{\newsymsec{reduction rule}{(T\protect\dash cs)}}
        \UnaryInfC{$(\Sigma_{i\le k}M_i)^+ := \Sigma_{i\le k}M_i^+$}
        \DisplayProof
      \end{center}
      \vspace{-0.5em}
      \caption{Contextual rules}
      \label{fig:fPCR}
    \end{subfigure}
  \caption{Full parallel reduction}\label{fig:def+}
\end{figure*}

\begin{lemma}
For any $M$ (resp. $Q$), $M^+$ (resp. $Q^+$) is well defined.
\end{lemma}
\begin{proof}
  By induction, since it is always the case that exactly  one rule is applied.
\end{proof}

\begin{lemma}\label{lemma:coflRta}
  If $M\Rta N$ (resp. $Q\Rta P$) then $N\Rta M^+$ (resp. $P\Rta Q^+$).
\end{lemma}
\begin{proof}
  By induction on $M$:
  \begin{itemize}
  \item If $M=x$:\\
    Then $N=x \Rta x=M^+$.
  \item If $M=\lambda x.M'$:\\
    Then $N=\lambda x.N'$ for some $N'$ such that $M'\Rta N'$.\\
    By IH, $N'\Rta M'^+$ and thus $N\Rta \lambda x.M'^+=M^+$.
  \item If $M= M_1\ M_2$:
    \begin{itemize}
    \item If $M_1$ is not an abstraction:\\
      Then $N=N_1\ N_2$ with $M_1\Rta N_1$ and $M_2\Rta N_2$.\\
      By IH, $N_1\Rta M_1^+$ and $N_2\Rta M_2^+$, thus $N\Rta M_1^+\ M_2^+=M^+$. 
    \item If $M_1=\lambda x.M_0$:
      \begin{itemize}
      \item Either $N=(\lambda x.N_0)\ N_2$ with $M_i\Rta N_i$ (for $i\in\{0,2\}$).\\
        By IH, $N_i\Rta M_i^+$ and $N\Rta M_0^+[M_2^+/x]=M^+$.
      \item Or $N= N_1[N_2/x]$ with $M_i\Rta N_i$ (for $i\in\{0,2\}$).\\
        By IH, $N_i\Rta M_i^+$ and $N\Rta M_0^+[M_2^+/x]=M^+$.
      \end{itemize}
    \item If $M_1=\Sigma_{i\in I}\bar\tau_{a_i\cons\alpha_i}(Q_{i})$:
      \begin{itemize}
      \item Either $N=(\Sigma_{i,j}\bar\tau_{a\cons\alpha_i}(P_{i,j}))\ N_2$  with $M_2\Rta N_2$ and $Q_{i}=\Sigma_iP'_{i,j}$ and $P'_{i,j}\Rta P_{i,j}$ .\\
        By IH, $N_2\Rta M_2^+$ and, moreover,\\ 
        $P_{i,j}\Rta Q_{i,j}^+=\Sigma_{k}Q'_{i,j,k}$ where $Q'_{i,j,k}$ that are not sums.\\ 
        Thus $N\Rta\Sigma_{i,j,k}\bar\tau_{\alpha_i}(Q'_{i,j,k}\pt\Pi_{\gamma\in a_i}\tau_\gamma(M_2^+)) = M^+$.
      \item Or $N=\Sigma_{i,j}\bar\tau_{\alpha_i}(P_{i,j}\pt\Pi_{\gamma\in a_i}\tau_\gamma(N_2))$ with $M_2\Rta N_2$ and $Q_{i}\Rta \Sigma_jP_{i,j}$.\\
        By IH, $N_2\Rta M_2^+$ and, moreover,\\ 
        $\Sigma_jP_{i,j}\Rta Q_{i}^+=\Sigma_{j,k}Q'_{i,j,k}$ where $Q'_{i,j,k}$ that are not sums and $P_{i,j}\Rta\Sigma_kQ'_{i,j,k} $.\\ 
        Thus $N\Rta\Sigma_{i,j,k}\bar\tau_{\alpha_i}(Q'_{i,j,k}\pt\Pi_{\gamma\in a_i}\tau_\gamma(M_2^+)) = M^+$.
      \end{itemize}
    \end{itemize}
  \item If $Q=\tau_\alpha(M)$:
    \begin{itemize}
    \item If $M$ is not an abstraction:\\
      Then $P=\tau_\alpha(N)$ for some $N$ such that $M\Rta N$.\\
      By IH, $N\Rta M^+$ and thus $P\Rta \lambda x.M^+=Q^+$.
    \item If $\alpha= a\cons\alpha$ and $M=\lambda x.M'$:
      \begin{itemize}
      \item Either $P=\tau_{a\cons\alpha}(\lambda x.N)$ with $M\Rta N$.\\
        By IH, $N\Rta M'^+$ and $P\Rta \tau_\alpha(M'^+[\bareps_a/x])=Q^+$.
      \item Or $P=\tau_\alpha(N[\bareps_a/x])$ with $M'\Rta N$.\\
        By IH, $N\Rta M'^+$ and $P\Rta \tau_\alpha(M'^+[\bareps_a/x])=Q^+$.
      \end{itemize}
    \item If $M=\Sigma_{i}\bar\tau_{\beta_i}(Q_i)$:
      \begin{itemize}
      \item Either $N=\tau_\alpha(\Sigma_{i,j}\bar\tau_{\beta_i}(P'_{i,j}))$ with $Q_i=\Sigma_jP_{i,j}$ and $P_{i,j}\Rta P'_{i,j}$.\\
        By IH, $P'_{i,j}\Rta P_{i,j}^+$. Thus, $N\Rta\Sigma_{\{i \mid \alpha\le\beta_i\}}\Sigma_jP_{i,j}^+=\Sigma_{\{i \mid \alpha\le\beta_i\}}Q_i^+=Q^+$.
      \item Or $N=\Sigma_{\{i\mid \alpha\le \beta_i\}}Q'_i$ with $Q_i\Rta Q_i'$.\\
        By IH, $Q_i'\Rta Q_i^+$. Thus, $N\Rta \Sigma_{i\mid \alpha\le\beta_i}Q_i^+=Q^+$.
      \end{itemize}
    \end{itemize}
  \item If $M=\Sigma_iM_i$:\\
    Then $N=\Sigma_i N_i$ with $M_i\Rta N_i$.\\
    By IH, $N_i\Rta M_i^+$ and $N\Rta \Sigma_iM_i^+ = M^+$.
  \item If $M=\bar\tau_{\alpha}(\Sigma_i Q_i)$ where none of the $Q_i$ are sums:\\
    Then we can only apply rules $(P\dash\bar\tau\+)$ and $(P\dash\pt\+)$. Thus there are $J$ and a surjective function $\phi:I\rta J$ such that $N=\Sigma_{j\in J}\bar\tau_{\alpha}(\Sigma_{i\in\phi^{-1}(j)} P_{i})$ and $Q_i \Rta P_i$.\\
    By IH, $P_i\Rta Q_i^+$ and $N\Rta\Sigma_{i\in I}\bar\tau_\alpha(Q_i^+)=M^+$.
  \item If $Q=\Pi_{i\le n}\Sigma_{j\le k_i} Q_{ij}$ where none of the $Q_{ij}$ are sums and where either $n\neq 1$ or one of the $k_i\neq 1$ :\\
      Then there are, for all $i\le n$, $J_i$ and $\phi_i:\llb 1,k_i\rrb\rta J_i$ such that $P=\Sigma_{(t_i)_i\in (J_i)_i}\Pi_{i\le n}\Sigma_{j\mid \phi(j)=t_i}P_{ij}$ with $Q_{ij}\Rta P_{ij}$.\\
      By IH, $P_{ij}\Rta Q_{ij}^+$ and $P\Rta\Sigma_{j_1\le k_1... j_n\le k_n}\Pi_{i\le n}Q_{ij_i}^+=Q^+$.
  \end{itemize}
\end{proof}

\begin{theorem}[Confluence] \label{th:confluence}
  The calculus $\Lam{D}$ with the reduction $\rta$ is confluent:
  \begin{alignat*}3
    \ M & \ \rta^* & \ M_2 \\
    \rotatebox[origin=c]{-90}{$\rta^{*}$}\ & \ \ \rotatebox[origin=c]{-45}{$\rightsquigarrow$} & \rotatebox[origin=c]{-90}{$\rta^*$}\ \\
    \ M_1 & \ \rta^* & \ M'
  \end{alignat*}
\end{theorem}
\begin{proof}
  By Lemma \ref{lemma:coflRta}, $\Rta$ is strongly confluent. This means that, for any $M_1\Leftarrow M\Rta M_2$, we have $M_1\Rta M^+\Leftarrow M_2$. By chasing diagrams, we obtain the confluence of $\Rta$ and we conclude by Lemma \ref{lemma:Rtarta} stating that $\Rta^*=\rta^*$.
\end{proof}

\subsubsection{Standardization theorem}\ \newline
This section is dedicated to the proof of Theorem~\ref{th:standardisation} stating a version of the standardization theorem for \Lam{D}. The proof is directly inspired by Kashima's proof \cite{Kas01}.

\begin{definition} 
  The \newdefsecprem{reduction}{standard}, denoted by \newsym{\protect\St} is defined in Figure \ref{fig:standardization}.
\end{definition}

\begin{figure*}
  \begin{center}
    \AxiomC{$M\rta^*_h x$}
    \RightLabel{\newsymsec{reduction rule}{(S\protect\dash x)}}
    \UnaryInfC{$M\St x$}
    \DisplayProof\hskip 60pt
    \AxiomC{$M\rta_h^* \lambda x.M_0$}
    \AxiomC{$M_0\St N_0$}
    \RightLabel{\newsymsec{reduction rule}{(S\protect\dash \lambda)}}
    \BinaryInfC{$M\St \lambda x.N_0$}
    \DisplayProof\\
    \vspace{1em}
    \AxiomC{$M\rta_h^* M_1\ M_2$}
    \AxiomC{$M_1\St N_1$}
    \AxiomC{$M_2\St N_2$}
    \RightLabel{\newsymsec{reduction rule}{(S\protect\dash \at)}}
    \TrinaryInfC{$M\St N_1\ N_2$}
    \DisplayProof\\
    \vspace{1em}
    \AxiomC{$P\rta_h^* \tau_\alpha(M)$}
    \AxiomC{$M\St N$}
    \RightLabel{\newsymsec{reduction rule}{(S\protect\dash \tau)}}
    \BinaryInfC{$P\St \tau_\alpha(N)$}
    \DisplayProof\hskip 20pt
    \AxiomC{$M\rta_h^* \Sigma_i\bar\tau_{\alpha_i}(P_i)$}
    \AxiomC{$\forall i,\ P_i\St Q_i$}
    \RightLabel{\newsymsec{reduction rule}{(S\protect\dash \protect\bar\tau)}}
    \BinaryInfC{$M\St \Sigma_i\bar\tau_{\alpha_i}(Q_i)$}
    \DisplayProof\\
    \vspace{1em}
    \AxiomC{$P\rta_h^* \Sigma_i P_i$}
    \AxiomC{$\forall i,\ P_i\St Q_i$}
    \RightLabel{\newsymsec{reduction rule}{(S\protect\dash +)}}
    \BinaryInfC{$P\St \Sigma_iQ_i$}
    \DisplayProof\hskip 30pt
    \AxiomC{$P\rta_h^* \Pi_i P_i$}
    \AxiomC{$\forall i,\ P_i\St Q_i$}
    \RightLabel{\newsymsec{reduction rule}{(S\protect\dash \pt)}}
    \BinaryInfC{$P\St \Pi_iQ_i$}
    \DisplayProof
  \end{center}
  \caption{Definition of the standard reduction}
  \label{fig:standardization}
\end{figure*}


\begin{prop}
  We have the following inclusions:
  \begin{itemize}
  \item $\St\ \subseteq\  \rta^*$,
  \item $id\subseteq \St$, {\em i.e.}, \St is reflexive,
  \item $\rta_h^*\ \subseteq\  \St$,
  \item $\St\ \subseteq\ \rta_h^*\rta_{\not h}^*$ where $\rta_{\not h}^*$ is the reflexive transitive closure of $\rta_{\not h}=\rta-\rta_{h}$.
  \end{itemize}
\end{prop}
\begin{proof}
  \begin{itemize}
  \item The inclusion $\St\ \subseteq\  \rta^*$ is obtain by easy induction (using each time the transitivity on $\rta_h^*\subseteq\rta^*$ and on the corresponding contextual rule of Figure \ref{fig:FCR} applied on the inductive hypothesis).
  \item The inclusion $id\subseteq \St$ derives from an easy induction using $id\subseteq\rta_h^*$.
  \item The inclusion $\rta_h^*\ \subseteq\  \St$ is obtained from a case analysis and the inclusion $id\subseteq \St$.
  \item Let $M,N\in\Lam D$ (resp. $P,Q\in\Test D$) be such that $M\St N$ (resp. $P\St Q$). We will show that $M\rta_h^*\rta_{\not h}^*N$ (resp. $P\rta_h^*\rta_{\not h}^*Q$) by induction on $N$ (resp. $Q$):
    \begin{itemize}
    \item If $N=x$ with $M\rta^*_h x$: trivial.
    \item If $N=\lambda x.N_0$, then $M\rta_h^* \lambda x.M_0$ and $M_0\St N_0$. By IH $M_0\rta_h^*\rta_{\not h}^*N_0$ so that Rule $(h\dash c\lambda)$ gives $M\rta_h^* \lambda x.M_0\rta_h^*\rta_{\not h}^*\lambda x.N_0$.
    \item If $N=N_1\ N_2$, then $M\rta_h^* M_1\ M_2$, $M_1\St N_1$ and $M_2\St N_2$. By induction hypothesis $M_1\rta_h^*M_1'\rta_{\not h}^*N_1$ for some $M_1'\in\Lam{D}$. 
      \begin{itemize}
      \item If $M_1'$ is not an abstraction, then there is no abstraction in the sequence \linebreak $M_1\rta_h\cdots\rta_h M_1'$ and by Rule $(h\dash c\at)$, $M\rta_h^* M_1\ M_2\rta_h^*M_1'\ M_2\rta_{\not h}^*N_1\ M_2$. 
      \item Otherwise, there is a first abstraction $M_1''$ such that $M_1\rta_h^* M_1''\rta^*M_1'$ with no abstraction in the sequence $M_1\rta_h\cdots\rta_h M_1''$.\\
        In this case, by Rule $(h\dash c\at)$,\\ 
        $M\rta_h^* M_1\ M_2\rta^*_h M_1''\ M_2\rta_{\not h}^*M_1'\ M_2\rta_{\not_h}^* N_1\ M_2\rta_{\not h}^* N_1\ N_2$.
      \end{itemize}
    \item If $Q=\tau_\alpha(N)$, then the argument is similar:\\
      There is $M$ such that $P\rta_h^*\tau_\alpha(M)$ and $M\St N$. By IH, there is $M'$ such that $M\rta_h^*M'\rta_{\not h}^*N$. Either $M'$ is not an abstraction and since there is no abstraction in the sequence $M\!\rta_h\!\cdots\!\rta_h\!M'$, we have, by Rule $(h\dash c\tau)$, that $P\!\rta_h\!\tau_\alpha(M){\rta^*_h}\tau_\alpha(M'){\rta^*_{\not h}}\tau_\alpha(N)$. Otherwise there is a first abstraction $M''$ in the sequence $M\rta_h\cdot\rta_h M''{\rta_h}\cdots{\rta_h} M'$, and we have, by Rule $(h\dash c\tau)$, that $P\rta_h\tau_\alpha(M)\rta^*_h\tau_\alpha(M'')\rta^*_{\not h}\tau_\alpha(N)$.
    \item If $N=\Sigma_i\bar\tau_{\alpha_i}(Q_i)$, there are $(P_i)_i$ such that $M\rta_h^* \Sigma_i\bar\tau_{\alpha_i}(P_i)$ and $P_i\St Q_i$ for all $i$. By IH, for all $i$, $P_i\rta_h^*P_i'\rta_{\not h}^*Q_i$ for some $P_i'\in\Lam{D}$. For all $i$, if $P_i'$ is not a sum (with $n\neq 1$ arguments) we set $P_i''=P_i'$, otherwise there is a first sum $P_i''$ such that $P_i\rta_h^* P_i''\rta_h^* P_i'$.\\ Then, using Rule $(h\dash c\bar\tau)$ we have, for all $i$, $\bar\tau_{\alpha_i}(P_i)\rta_h^*\bar\tau_{\alpha_i}(P_i'')\rta_{\not h}^*\bar\tau_{\alpha_i}(Q_i)$.\\ Thus, using Rule $(h\dash cs)$, we have $M\rta_h^* \Sigma_i\bar\tau_{\alpha_i}(P_i)\rta_h^* \Sigma_i\bar\tau_{\alpha_i}(P''_i)\rta_{\not h}^* \Sigma_i\bar\tau_{\alpha_i}(Q_i)$.
    \item If $Q=\Pi_i(Q_i)$ then the argument is similar:\\
      There are $(P_i)_i$ such that $P\rta_h^* \Pi_iP_i$ and $P_i\St Q_i$ for all $i$. By IH, for all $i$,\linebreak $P_i\rta_h^*P_i'\rta_{\not h}^*Q_i$ for some $P_i'\in\Lam{D}$. For all $i$, if $P_i'$ is not a sum (with $n\neq 1$ arguments) we set $P_i''=P_i'$, otherwise there is a first sum $P_i''$ such that $P_i\rta_h^* P_i''\rta_h^* P_i'$.\\ Then, using Rule $(h\dash c\pt)$, we have $P\rta_h^* \Pi_iP_i\rta_h^* \Sigma_iP''_i\rta_{\not h}^* \Sigma_iQ_i$.
    \item If $Q=\Sigma_i(Q_i)$, there are $(P_i)_i$ such that $P\rta_h^* \Sigma_iP_i$ and $P_i\St Q_i$ for all $i$. By IH, for all $i$, $P_i\rta^*_hP_i'\rta^*_{\not h} Q_i$ and, by Rule $(h\dash \pt)$, $\Sigma_iP_i\rta^*_h\Sigma_iP_i'\rta^*_{\not h} \Sigma_iQ_i$.
    \end{itemize}
  \end{itemize}
\end{proof}


\begin{lemma}\label{lemma:standBartau+}
  Ultimately, sums will necessarily commutes with $\bar\tau$, with products and with $\tau$:
  \begin{enumerate}
  \item If $P\rta_h^* \Sigma_{j\le k} Q_j$, then there for all $j\le k$ is $P_j\rta^*_h Q_j$ such that 
    $$\bar\tau_\alpha(P)\rta_h^*\Sigma_{j\le k} \bar\tau_\alpha(P_j).$$
  \item Similarly, if $P\rta_h^* \Sigma_{j\le k} Q_{j}$, then for all $j\le k$, there is $P_{j}\rta^*_h Q_{j}$ such that 
    $$Q\pt P\rta_h^*\Sigma_{j}(Q\pt P_{j}).$$ 
  \item Similarly, if $M\rta_h^* \Sigma_{j\le k} \bar\tau_{\beta_j}(Q_j)$, then for all $j\le k$, there is $P_j\rta^*_h Q_j$ such that 
    $$\tau_\alpha(M)\rta_h^*\Sigma_{\{j\mid \beta_j\ge \alpha\}} P_j.$$
  \end{enumerate}
\end{lemma}
\begin{proof}
  The proof follows the exact same pattern for each cases.
  \begin{enumerate}
  \item Let $P\rta_h^n \Sigma_{j\le k} Q_j$. The proof is by induction on the lexicographically ordered $(n,P)$.
    \begin{itemize}
    \item If $n=0$ then this is Rule $(\bar\tau+)$.
    \item If $P=\Sigma_{i\le k'}P_i'$ with $k'\neq 1$, there is a surjective $\phi: [1,k]\rta [1,k']$ such that $P'_i\rta_h^{n_i}\Sigma_{j\in\phi^{-1}(i)}Q_j$ with $n=\Sigma_i n_i$. By IH on each $P_i'$, there are $(P_j)_{j\in\phi^{-1}(i)}$ such that, for all $i\le k'$, $\bar\tau_\alpha(P'_i)\rta_h^*\Sigma_{j\in\phi^{-1}(i)} \bar\tau_\alpha(P_j)$ with $P_j\rta^*_h Q_j$. Thus $\bar\tau_\alpha(P)\Ruledarrow{\bar\tau+}{h}{}\Sigma_{i\le k'}\bar\tau_\alpha(P'_i)\rta^*_h\Sigma_{i\le k'}\Sigma_{j\in\phi^{-1}(i)}\bar\tau_\alpha(P_j)$.  
    \item Otherwise, we can decompose the reduction by $P\rta_hP'\rta_h^{n-1}\Sigma_{j\le k} Q_j$. Since $P$ is not a sum we can apply the rule $H\dash c\bar\tau$ so that $\bar\tau_\alpha(P)\rta_h\bar\tau_\alpha(P')$ and we conclude since by IH, $\bar\tau_\alpha(P')\rta_h^*\Sigma_{j\le k} \bar\tau_\alpha(P_j)$.
    \end{itemize}
  \item Let $P\rta_h^n \Sigma_{j\le k} Q_j$. The proof is by induction on the lexicographically ordered $(n,P)$.
    \begin{itemize}
    \item If $n=0$ then this is Rule $(\pt+)$.
    \item If $P=\Sigma_{i\le k'}P_i'$ with $k'\neq 1$, there is a surjective $\phi: [1,k]\rta [1,k']$ such that $P'_i\rta_h^{n_i}\Sigma_{j\in\phi^{-1}(i)}Q_j$ with $n=\Sigma_i n_i$. By IH on each $P_i'$, there are $(P_j)_{j\in \phi^{-1}(i)}$ such that, for all $i$, $(Q\pt P_i)\rta_h^*\Sigma_{j\in\phi^{-1}(i)} (Q\pt P_j)$ with $P_j\rta^*_h Q_j$. Thus $Q\pt P\Ruledarrow{\pt+}{h}{}\Sigma_{i\le k'}(Q\pt P_i)\rta^*_h\Sigma_{i\le k'}\Sigma_{j\in\phi^{-1}(i)}(Q\pt P_j)$.  
    \item Otherwise, we can decompose the reduction by $P\rta_hP'\rta_h^{n-1}\Sigma_{j\le k} Q_j$. Since $P$ is not a sum we can apply the rule $H\dash c\pt$ so that $Q\pt P\rta_hQ\pt P'$ and we conclude since by IH, $Q\pt P'\rta_h^*\Sigma_{j\le k} Q\pt P_j$.
    \end{itemize}
  \item Let $M\rta_h^n \Sigma_{j\le k} \bar\tau_\alpha(Q_j)$. The proof is by induction on the lexicographically ordered $(n,M)$:
    \begin{itemize}
    \item If $n=0$ then this is Rule $(\tau\bar\tau)$.
    \item If $M=\Sigma_{i\le k'}\bar\tau_{\gamma_i}(P_i')$ with $k'\neq 1$, there is a surjective $\phi: [1,k]\rta [1,k']$ such that $\bar\tau_{\gamma_i}(P'_i)\rta_h^{n_i}\Sigma_{j\in\phi^{-1}(i)}\bar\tau_{\beta_j}Q_j$ with $n=\Sigma_i n_i$. By IH on each $\bar\tau_{\gamma_i}(P_i')$, there are $(P_j)_{j\le \phi^{-1}(i)}$ such that, for all $i$, $\tau_\alpha(\bar\tau_{\gamma_i}(P'_i))\rta_h^*\Sigma_{\{j\in\phi^{-1}(i)\mid \alpha\le\beta_j\}} P_j$ with $P_j\rta^*_h Q_j$. Since the only head reduction that can be applied on each $\tau_\alpha(\bar\tau_{\gamma_i}(P'_i))$ is $(h\dash \tau\bar\tau)$, we have that $\tau_\alpha(M)\rta_h\Sigma_{\{\i\mid \alpha\le\gamma_i\}}P_i\rta_h^*\Sigma_j Q_j$.
    \item The case $M=\lambda x.M'$ is impossible since $M\rta^*\Sigma_j\bar\tau_{\beta_j}(Q_j)$ and no rule can erase a $\lambda$ in first position.
    \item Otherwise, we can decompose the reduction by $M\rta_hM'\rta_h^{n-1}\Sigma_{j\le k} \bar\tau_{\beta_j}(Q_j)$. Since $M$ is not an abstraction we can apply the rule $(h\dash \tau)$ so that $\tau_\alpha(M)\rta_h\tau_\alpha(M')$ and we conclude since by IH, $\tau_\alpha(M')\rta_h^*\Sigma_{\{j\mid \beta_j\ge\alpha\}} P_j$.
    \end{itemize}
  \end{enumerate}
\end{proof}

\begin{lemma}\label{lemma:standardisation}
  For all $M,N,N'\in \Lam{D}$ such that $M\St N\rta N'$, there is $M'$ such that $M\St N'$.\\
  Similarly, for all $P,Q,Q'\in\Test{D}$ such that $P\St Q\rta Q'$, there is $P'$ such that $P\St Q'$.
\end{lemma}
\begin{proof}
  We proceed by structural induction on $N$: \todo{do the induction on $N\rta N'$}
  \begin{itemize}
  \item The case $N=x$ is impossible since $x$ is a normal form.
  \item If $N=\lambda x. N_0$ then $N_0\rta N_0'$ with $N'=\lambda x.N'_0$. By definition of $\St$, $M\rta_h^*\lambda x.M_0$ and $M_0\St N_0$. By IH, $M_0\St N_0'$, thus $M\St \lambda x.N_0'$.
  \item The case $N=\0$ is impossible since $\0$ is a normal form.
  \item If $N=\bar\tau_{\alpha}(Q)$ then the only rule that can change the form of the expression is $(\bar\tau\+)$ applied in head position:
    \begin{itemize}
    \item Either $N=\bar\tau_{\alpha}(\Sigma_jQ_j)\Ruledarrow{\bar\tau+}{h}{}N'=\Sigma_j\bar\tau_{\alpha}(Q_j)$. By definition of $\St$, $M\rta_h^* \bar\tau_{\alpha}(P)$ and $P\rta_h^*\Sigma_jP_j$ with $P_j\St Q_j$. Thus, by Lemma \ref{lemma:standBartau+}, there is $(P'_j)_j$ such that $M\rta_h^*\Sigma_j\bar\tau_\alpha(P'_j)$ with $P'_j\rta^*P_j\St Q_j$, so that $M\St N'$.
    \item Otherwise, $Q\rta Q'$ and $N'=\bar\tau_\alpha(Q')$. In this case, since $M\rta^*_h\bar\tau_\alpha(P)$ and $P\St Q\rta Q'$, we can apply the IH so that $P\St Q'$ and $M\St\bar\tau_\alpha(Q')$.
    \end{itemize}
  \item Let $N=\Sigma_{i\le n}N_i$ with $n>0$. Then, modulo commutativity of the sum, we can assume that $N_n\rta N_n'$, so that $N'=\Sigma_{i<n}N_i+ N'_n$. By definition of $\St$, $M\rta_h^*\Sigma_{i\le n}M_i$ with $M_i\St N_i$. By induction hypothesis, $M_n\St N'_n$ and we can set $M\St N'$.
  \item Let $N=N_1\ N_2$, then $M\rta_h^*M_1\ M_2$ with $M_1\St  N_1$ and $M_2\St  N_2$. There are different cases:
    \begin{itemize}
    \item Either $N_1\rta_h N_1'$ and $N'=N_1'\ N_2$. In this case, the IH on $M_1\St N_1\rta_h N_1'$ gives $M_1\St N_1'$, so that $M\St N'$.
    \item Or $N_2\rta_h N_2'$ and $N'=N_1\ N_2'$. In this case, the IH on $M_2\St N_2\rta_h N_2'$ gives $M_2\St N_2'$, so that $M\St N'$.
    \item Or $N_1 = \lambda x. N_0$ and $N'=N_0[N_2/x]$. By definition of $\St$, $M_1\rta_h^*\lambda x.M_0$ with $M_0\St N_0$. By easy induction on $\St$, one can see \todo{write lemma but easy} that $M_0[M_2/x]\St N_0[N_2/x]$. We can conclude since $\rta^*\St\subseteq\St$. 
    \item Or $N_1 = \Sigma_{i\le n}\bar\tau_{a_i\cons \alpha_i}(Q_i)$ and $N'=\Sigma_{i\le n}\bar\tau_{\alpha_i}(Q_i\pt\Pi_{\gamma\in a_i}\tau_{\gamma}(N_2))$. By definition of $\St$, $M_1\rta_h^*\Sigma_{i\le n}\bar\tau_{a_i\cons \alpha_i}(P_i)$ and $P_i\St Q_i$ for all $i$. By definition of $\St$, one can see \todo{write lemma but easy} that \linebreak $\Sigma_{i\le n}\bar\tau_{\alpha_i}(P_i\pt\Pi_{\gamma\in a_i}\tau_{\gamma}(M_2))\St\Sigma_{i\le n}\bar\tau_{\alpha_i}(\Pi_{\gamma\in a_i}\tau_{\gamma}(N_2))$ so that $M\St N$. 
    \end{itemize}
  \item If $Q = \tau_{a\cons\alpha}(N)$, then $P\rta_h^*\tau_{a\cons\alpha}(M)$ with $M\St N$ and there are different cases:
    \begin{itemize}
    \item Either $N\rta N'$ and $Q' = \tau_{a\cons\alpha}(N')$. In this case, the IH on $M\St N\rta N'$ gives $M\St N'$, so that $P\St Q'$.
    \item Or $N = \lambda x. N_0$ and $Q' = \tau_{\alpha}(N_0[\bareps_a/x])$. By definition of $\St$, $M\rta_h^*\lambda x.M_0$ with $M_0\St N_0$. By easy induction on $\St$, one can see \todo{write lemma but easy} that $M_0[\bareps_a/x]\St N_0[\bareps_a/x]$. We can conclude since $\rta^*\St\subseteq\St$. 
    \item Or $N = \Sigma_{i\le n}\bar\tau_{\beta_i}(Q_i)$ and $N'=\Sigma_{i\le n \mid \beta_i\ge \alpha}Q_i$. By definition of $\St$, $M\rta_h^*\Sigma_{i\le n}\bar\tau_{\beta_i}(P_i)$ and $P_i\St Q_i$ for all $i$.  By Lemma \ref{lemma:standBartau+}, there is $(P_i')_i$ such that $\tau_\alpha(M)\rta_h^*\Sigma_{i\le n \mid \beta_i\ge \alpha}P'_i$ and $P_i'\St Q_i$ so that $P\St Q'$.
    \end{itemize}
  \item If $Q = \Sigma_{i\le n} Q_i$ then (up to commutativity of the sum) $Q_n\rta Q_n'$ and $Q'=\Sigma_{i< n} Q_i+Q_n'$. By definition of $\St$, $P\rta^*_h \Sigma_{i\le n} P_i$ with $P_i\St Q_i$ for all $i$. By IH on $P_n\St Q_n\rta Q_n'$, $P_n\St Q'_n$ so that $P\St Q'$.
  \item If $Q = \Pi_{i\le n} Q_i$ then the only rule that changes the form of the expression is $(\pt\+)$ applied in head position. There are two cases:
    \begin{itemize}
    \item Either $Q=\Pi_i\Sigma_{j\le k_i}Q_{ij}\Ruledarrow{\pt\+}{h}{}Q'=\Sigma_{(j_i)_i}\Pi_i Q_{ij_i}$. By definition of $\St$ (used 2 times), $P\rta_h^* \Pi_i P_i$ and $P_i\rta_h^*\Sigma_{j\le k_i}P_{ij}$ with $P_{ij}\St Q_{ij}$ for all $i,j$. Thus, by Lemma \ref{lemma:standBartau+}, there is $(P'_{ij})_{ij}$ such that $P\rta_h^*\Sigma_{(j_i)_i}\Pi_i P'_{ij}$ with $P'_{ij}\rta^*P_{ij}\St Q_{ij}$, so that $M\St N'$.
    \item Otherwise (and up to commutativity of the sum), $Q_n\rta Q_n'$ and $Q'=\Pi_{i< n}Q_i \pt Q_i'$. By definition of $\St$, $P\rta^*_h \Pi_i P_i$ and $P_i\St Q_i$. We can apply the IH on $P_n\St Q_n\rta_h Q_n'$, so that $P\St Q'$.
    \end{itemize}
  \end{itemize}
\end{proof}

\begin{theorem}[Standardization]\label{th:standardisation}
  For any reduction $M\rta^* N$ (resp. $P\rta^* Q$), there is a standard reduction $M\St N$ (resp. $P\St Q$).
  In particular, any term $M$ (resp. test $Q$) head converges iff it reduces to a may head-normal form:
  \begin{align*}
    M\Da\quad &\Lra\quad\exists N\in mhnf, M\rta^*N'  &  P\Da\quad &\Lra\quad\exists Q\in mhnf, P\rta^*Q'. 
  \end{align*}
\end{theorem}
\begin{proof}
  By applying successively Lemma~\ref{lemma:standardisation}. The equivalence between $\Da$ and having a may-head-normal form is an immediate consequence once noticed that whenever $M\rta_{\not h} M'$ then $M\in\mhnf$ iff $M'\in mhnf$.
\end{proof}

\subsection{Invariance for the convergence}\ \newline
We will see in this section that the head convergence in at most $n$ steps is invariant wrt the reduction. This means that performing a non-head reduction can only reduce the length of convergence.

\begin{theorem}[Invariance for the convergence]\label{th:invHeadReducibility}
  For any terms $M\rta N$ (resp. test $P\rta Q$) and any $n\in \Nat$:
  \begin{align*}
    M\Da_{\!\!\! n}\quad &\Rta\quad N\Da_{\!\!\! n}       &       P\Da_{\!\!\! n}\quad &\Rta\quad Q\Da_{\!\!\! n} 
  \end{align*}
\end{theorem}
\begin{proof}
  By recursive invocations of Lemma~\ref{confH}, for any $k$ we can close the diagrams:
  \begin{alignat*}7
    \ M & \ \rta_h & \ M_1 & \hspace{10em} & \ Q & \ \rta_h & \ Q_1\\
    \rotatebox[origin=c]{-90}{$\rta$}_k\ &\ \ \rotatebox[origin=c]{-45}{$\rightsquigarrow$}  & \rotatebox[origin=c]{-90}{$\rta^*$}\ & & \rotatebox[origin=c]{-90}{$\rta$}_k\ &\ \ \rotatebox[origin=c]{-45}{$\rightsquigarrow$} & \rotatebox[origin=c]{-90}{$\rta^*$}\ \\
    \ M_2 & \ \rta_h^? & \ M' & & \ Q_2 & \ \rta_h^? & \ Q'
  \end{alignat*}
  where $\rta_h^?:=(\rta_h\cup id)$ is either a head reduction or an equality.\\
  Recursively invoking this diagrams, for any $n$ we can now close the diagrams:
  \begin{alignat*}7
    \ M & \ \rta^n_h & \ M_1 & \hspace{10em} & \ Q & \ \rta_h & \ Q_1\\
    \rotatebox[origin=c]{-90}{$\rta^*$}\ &\ \ \rotatebox[origin=c]{-45}{$\rightsquigarrow$}  & \rotatebox[origin=c]{-90}{$\rta^*$}\ & & \rotatebox[origin=c]{-90}{$\rta^*$}\ &\ \ \rotatebox[origin=c]{-45}{$\rightsquigarrow$} & \rotatebox[origin=c]{-90}{$\rta^*$}\ \\
    \ M_2 & \ \rta_h^{\le n} & \ M' & & \ Q_2 & \ \rta_h^{\le n} & \ Q'
  \end{alignat*}
  where $\rta_h^{\le n}=\bigcup_{i\le n}\rta_h^i$ represents at most $n$ iterations of $\rta$.\\
  In particular, if $M\rta^*_h M'$ with $M'\in mhnf$ ({\em i.e.} $M$ converges), since $M\rta N$, there is $N_0$ such that $N\rta_h^{\le n}N_0$ and $N\rta^* N_0$, from the last we deduce that $N_0\in mhnf$ and conclude. The same goes for tests.
\end{proof}

\todo{see comments}

In order to prove this theorem we need a stronger notion of confluence for the cases where one of the reduction is a head reduction.

\begin{lemma}\label{confH}
 Any pick, $M\rta_hM_1$ and $M\rta M_2$ (resp. $Q\rta_hQ_1$ and $Q\rta Q_2$), between a head reduction and any reduction verifies the diamond:
  \begin{alignat*}7
    \ M & \ \rta_h & \ M_1 & \hspace{10em} & \ Q & \ \rta_h & \ Q_1\\
    \rotatebox[origin=c]{-90}{$\rta$}\ &\ \ \rotatebox[origin=c]{-45}{$\rightsquigarrow$}  & \rotatebox[origin=c]{-90}{$\rta^*$}\ & & \rotatebox[origin=c]{-90}{$\rta$}\ &\ \ \rotatebox[origin=c]{-45}{$\rightsquigarrow$} & \rotatebox[origin=c]{-90}{$\rta^*$}\ \\
    \ M_2 & \ \rta_h^? & \ M' & & \ Q_2 & \ \rta_h^? & \ Q'
    \end{alignat*}
    where $\rta_h^?:=(\rta_h\cup id)$ is either a head reduction or an equality.
\end{lemma}
\begin{proof}
   By induction on $M$ and $Q$:
  \begin{itemize}
  \item The cases $M=x$ and $M=\0$ are impossible since $M\rta_h M_1$.
  \item If $M=\lambda x.N$: then $M_1=\lambda x.N_1$ and $M_2=\lambda x.N_2$ so that $N_1\mathop{_h\!\!\leftarrow}N\rta N_2$, thus, by induction, there is $N'$ such that $N_1\rta^*N'\mathop{^?_h\!\!\leftarrow}N_2$, finally we can choose $M'=\lambda x.N'$.
  \item If $M= \Sigma_{i\le n+2}N^i$: then, modulo commutativity of the sum, $M_1=N_1^{n+2}\+\Sigma_{i\le n+1}N^i$ with $N^{n+2}\rta_hN_1^{n+2}$.
    \begin{itemize}
    \item Either (modulo commutativity of the sum), $M_2=N_2^{n+2}\+\Sigma_{i\le n+1}N^i$ with $N^{n+2}\rta N_2^{n+2}$ and by induction there is $N_1^{n+2}\rta^*N_0^{n+2}\mathop{^?_h\!\!\leftarrow}N_2^{n+2}$ such that $M'=N_0^{n+2}\+\Sigma_{i\le n+1}N^i$.
    \item Or (modulo commutativity of the sum), $M_2=N^{n+2}\+N_2^{n+1}\+\Sigma_{i\le n+1}N^i$ with \linebreak $N^{n+1}\rta N_1^{n+1}$,  so that $M'=N_1^{n+2}\+N_2^{n+1}\+\Sigma_{i\le n+1}N^i$.
    \end{itemize}
  \item If $M= \bar\tau_{\alpha}(Q)$ with $Q$ that is not a sum: then $M_1=\bar\tau_{\alpha_i}(Q_1)$ and $M_2=\bar\tau_{\alpha}(Q_2)$ with $Q_1\mathop{_h\!\!\leftarrow}Q\rta Q_2$, thus, by induction, there is $Q'$ such that $Q_1\rta^*Q'\mathop{^?_h\!\!\leftarrow}Q_2$, finally we can fix $M'=\bar\tau_{\alpha}(Q')$. 
  \item If $M= \bar\tau_{\alpha}(\Sigma_{i\le n+1} Q^i)$ and $M_1=\Sigma_{i\le n+1}\bar\tau_{\alpha}(Q^i)$:
    \begin{itemize}
    \item Either $M_2=\bar\tau_{\alpha}(Q_2^{n+1}\Sigma_{i\le n}Q^i)$ and $M'=\bar\tau_{\alpha}(Q^{n+1})\Sigma_{i\le n}\bar\tau_{\alpha}(Q^i)$.
    \item Or $Q^i = \Sigma_{j}P^{i,j}$ and $M_2=\Sigma_{j}\bar\tau_{\alpha}(P^{i,j})$, then $M'=\Sigma_{i,j}\bar\tau_{\alpha}(P^{i,j})$.
    \end{itemize}
  \item If $M= N\ L$:
    \begin{itemize}
    \item If $N$ is not an abstraction: then $M_1=N_1\ L$ with $N\rta_h N_1$. Moreover
      \begin{itemize}
      \item Either $M_2=N_2\ L$ with $N\rta N_2$ and $N_2$ that is not an abstraction. By induction there is $N'$ such that $N_1\rta^*N'{^?_h\leftarrow}N_2$, and $M'=N'\ L$.
      \item Or $M_2=(\lambda x.N_2)\ L$ with $N\rta N_2$ and $N_2$ that is an abstraction: since $N$ is not an abstraction,  this can only be the result of a $(\beta)$ or a $\bar\tau$ reduction in outermost position in $N$. In both cases, necessary $M_1=M_2$.
      \item Or $M_2=N\ L_2$ with $L\rta L_2$: then $M'= N_1\ L_2$.
      \end{itemize}
    \item If $N=\lambda x.N'$ : then $M_1=N'[L/x]$ and 
      \begin{itemize}
      \item Either $M'=M_2=M_1$.
      \item Or $M_2=(\lambda x.N'_2)\ L$ with $N'\rta N_2$, thus $M'=N_2'[L/x]$.
      \end{itemize}
    \item If $N=\Sigma_i\bar\tau_{\alpha_i}(Q_i)$: idem.
    \end{itemize}
  \item If $Q= \tau_\alpha(M)$:
    \begin{itemize}
    \item If $M$ is not an abstraction: then $Q_1=\tau_\alpha(M_1)$ and $Q_2=\tau_\alpha(M_2)$ with $M_1\mathop{_h\!\!\leftarrow}M\rta M_2$ and by induction hypothesis, there is $M'$ so that  $M_1\rta^*M'{^?_h\leftarrow}M_2$.
      \begin{itemize}
      \item Either $M_2$ is not an abstraction and $Q'=\tau_\alpha(M')$.
      \item Or $M\rta M_2$ is an abstraction created by a $(\beta)$ or a $(\bar\tau)$ outermost reduction. In both cases, necessary $M_1=M_2$.
      \end{itemize}
    \item If $M=\lambda x.N$: then $Q_1=\tau_{\alpha'}(N[\bareps_a/x])$ and 
      \begin{itemize}
      \item Either $Q_2=Q_1=Q'$.
      \item Or $Q_2=\tau_{\alpha}\lambda x.N_2$ with $N\rta N_2$, thus $Q'=\tau_{\alpha'}(N_2[\bareps_a/x])$.
      \end{itemize}
    \item If $M= \Sigma_{i\le n+1}\bar\tau_{\beta_i}(P^i)$: then $Q_1=\Sigma_{\{i\le n+1\mid \alpha\le \beta_i\}}P^i$ and 
      \begin{itemize}
      \item Either $Q_2=Q_1=Q'$.
      \item Or $Q_2= \tau_{\alpha}\Bigl(\Sigma_{i\le n}\bar\tau_{\alpha_i}(P^i) + \Sigma_j\bar\tau_{\beta_n}(R^j)\Bigr) $ with $\bar\tau_{\alpha_{n+1}}(P^{n+1})\rta \Sigma_j\bar\tau_{\beta_j}(R^j)$,\\ thus $Q'=\Sigma_{\{j\mid \alpha\le \beta_n\}}R^j+\Sigma_{\{i\le n\mid \alpha\le\beta_i\}}P^i$.
      \end{itemize}
    \end{itemize}
  \item If $Q= P\+ R$: then, modulo commutativity of the sum, $Q_1= P_1\+R$ with $P\rta_h P_1$.
   \begin{itemize}
   \item Either $Q_2 = P_2\+R$ with $P\rta P_2$ and the induction hypothesis gives $P'$ so that $M'= P'\+R$.
   \item Or $Q_2 = P\+R_2$ and $M'=P_1\+R_2$.
   \end{itemize}
  \item If $Q= P\pt R$: same as for $Q= P\+ R$ except if a rule $(\pt+)$ is used in outermost position. In this case, either only one of the reduction is a $(\pt+)$ and the two reductions are independents, or both of them are $(\pt+)$, which is similar to the case $M= \bar\tau_{\alpha}(\Sigma_{i\le n+1} Q^i)$.
  \end{itemize}
\end{proof}

\section{Proof}
  \subsection{Hyperimmunity implies full abstraction} \quad \newline
    \label{ssec:HtoFA}
    \label{ssec:HtoFA2}
%

In this subsection we show that if $D$ is sensible for \Lam D and is hyperimmune, $D$ is inequationally fully abstract for $\Lamb$, that is Theorem~\ref{th:FA}. We use the full abstraction of $D$ for \Lam{D} of Theorem~\ref{th:FAwT} (or rather its technical counterpart: Theorem~\ref{th:caracFAwT}) in order to express the problem in a purely syntactical form:
\begin{align*}
  \llb M\rrb \neq \llb N\rrb \quad 
  &\Longleftrightarrow\quad {\exists\alpha\in P},\quad\ \alpha \in\llb M\rrb -\llb N\rrb &\text{or conv.}\\
  &\stackrel{(1)}{\Longleftrightarrow}\quad  {\exists\alpha\in P},\quad\ \tau_\alpha(M){\Downarrow} \text{ and } \tau_\alpha(N){\Uparrow} &\text{ or conv.}\\
  &\stackrel{(2)}{\;\Longrightarrow}\quad {\exists C\in \Lambda^{\llc.\rrc},}\ C\llc M\rrc{\Downarrow} \text{ and } C\llc N\rrc {\Uparrow}&\text{or conv.}\\
  &\Longleftrightarrow\quad \phantom{\exists C\in \Lambda^{\llc.\rrc},}\ M\not\equiv_{\Hst} N
\end{align*}
Here $(1)$ is given by Theorem~\ref{th:caracFAwT} so that we only have to prove $(2)$ which is done in the proof of Theorem~\ref{th:FAtecn} by induction on the finite reduction $\tau_\alpha(M){\Downarrow}$. However, the proof require a specific treatment of the case where $M=\I$ (we have some $\eta\infty$-ex pensions issues) this is the purpose of the key-lemma (Lemma~\ref{lemma:fonda}). This key-lemma is assuming that $(2)$ is false for $M=\I$ (and any $N$) then co-inductively constructs a counterexample $(\alpha_n)_n$ the hyperimmunity by unfolding $\tau_\alpha(N){\Uparrow}$.

Before that, we need the technical Lemma~\ref{lemma:BTlemmes} in order to refute the operational equivalence between two $\lambda$-terms in easy cases.

\begin{lemma}[\cite{Wad76}] \label{lemma:BTlemmes}
  Let $M=\lambda x_1...x_n.y\ M_1\cdots M_k\in\Lamb$ and let $N=\lambda x_1...x_{n'}.y'\ N_1\cdots N_{k'}\in\Lamb$ be $\lambda$-terms such that $M\leob N$. Then:
  \begin{enumerate}
  \item $y=y'$, \label{eq:y=y'}
  \item $n-k=n'-k'$, \label{eq:m-n}
  \item if $i\le k$ and $i\le k'$ then $M_i\leob N_i$, \label{eq:MeqN}
  \item if $i> k$ and $i\le k'$ then $x_{i-k}\leob  N_i$, \label{eq:Meqx}
  \item if $i\le k$ and $i> k'$ then $M_{i-k}\leob  x_i$. \label{eq:Meqx2}
  \end{enumerate}
\end{lemma}
\begin{proof}
  From each $i\le 5$, assuming statements (1)...(i-1) and refuting statement (i), we can exhibit a context $C\in\Lcont$ such that $C\llb M\rrb\Da$ and $C\llb N\rrb\Ua$.
\end{proof}

\subsubsection{The key-lemma}\label{sssec:Crit}\ \newline
From now on, we consider an extensional \Kweb $D$ that is hyperimmune and sensible for $\Lam{D}$.\linebreak
The following lemma is a key lemma that introduces the hyperimmunity in the picture. It basically states that if $\tau_\alpha(N[\bareps_\alpha/x_0])\Ua$ then $N\ngeob x_0$.

\begin{lemma}\label{lemma:fonda}
  Let $\alpha\in D$ and $a_0,\dots,a_k\in\Achf{D}$ be such that $\alpha\in a_0$.\\
  Let $N\in\Lamb$ and $x_0,\dots,x_k$ be such that $\tau_\alpha(N[\bareps_{a_0}/x_0,\dots,\bareps_{a_k}/x_k])\Ua$. Then $N\ngeob x_0$.
\end{lemma}
\begin{proof}
  We define the recursive function $g_{N'}$ for any $N'\in\Lamb$ such that $N'\geob x_0$, it is done by recursively defining $g_{N'}(k)$ for $k\in \Nat$:\\
  Since $N'\geob x_0$, $N'$ is converging, and by Lemma~\ref{lemma:BTlemmes} $N'\rta_h^* \lambda y_1...y_n.x_0\ N_1\cdots N_n$ with $N_m\geob y_m$ for all $m\le n$. We then define $g_{N'}(0)=n$ and $g_{N'}(k+1)=max_{i\le n}g_{N_i}(k)$. \\
  We will show that assuming $N\geob x_0$ contradicts the hyperimmunity of $D$ by showing that: \medskip 

  There exists $(\alpha_n)_{n\ge 0}$ with {$\alpha_0=\alpha$} and for all $n$, $\alpha_n=a_1^n\cons\cdots a_{g_N(n)}^n\cons\alpha_n'$ and $\alpha_{n+1}\in\bigcup_{i\le g_N(n)}a_i^n$.\medskip\\
  We are constructing $(\alpha_n)_n$ by co-induction.\\
  Since $N\geob x_0$, it is converging, and by Lemma~\ref{lemma:BTlemmes}, $N\rta^* \lambda y_1...y_n.x_0\ N_1\cdots N_n$ with $N_m\geob y_m$ for all $m\le n$.\\
  We will assume that $\alpha=b_1\cons\cdots\cons b_n\cons\alpha'$ and $ a_0=\{\alpha,\beta_1,\dots,\beta_t\}$ with $\beta_i=c_1^i\cons\cdots\cons c_n^i\cons \beta_i'$ (always possible since ``$\cons$'' is a bijection).\\  
  Then (notice the use of a calculation done in Example~\ref{ex:barepsM1...Mn})
  \begin{align*}
    \tau_\alpha(N[s]) & \rta^* \tau_\alpha(\lambda y_1...y_n.\bareps_{a_0}\ N_1[s]\cdots N_n[s]) \\
    &\ruledarrow{\tau}{h}{*} \tau_{\alpha'}(\bareps_{a_0}\ N_1[s,s']\cdots N_n[s,s'])\\
    &\Ruledarrow{Ex\ref{ex:barepsM1...Mn}}{}{*} \tau_{\alpha'}(\Sigma_{d_1\cons\cdots d_n\cons\delta\in a_0}\bar\tau_{\delta}(\Pi_{m\le n}\Pi_{\gamma\in d_m}\tau_\gamma(N_m[s,s']))) \\
    &\ruledarrow{\tau\bar\tau}{h}{} \Pi_{m\le n}\Pi_{\gamma\in b_m}\tau_\gamma(N_m[s,s'])\quad +\quad \Sigma_{\{i\le t\mid \alpha'\le\beta'_i\}}\Pi_{m\le n}\Pi_{\gamma\in c_m^i}\tau_\gamma(N_m[s,s'])
  \end{align*}
  with $[s]=[\bareps_{a_0}/x_0,\dots,\bareps_{a_k}/x_k]$ and $[s']=[\bareps_{b_1}/y_1,\dots,\bareps_{b_n}/y_n]$.\\
  Since $\tau_\alpha(N[s])$ diverges, by standardization theorem (Th.~\ref{th:standardisation}), the test $\Pi_{m\le n}\Pi_{\gamma\in b_m}\tau_\gamma(N_m[s,s'])$ diverges. In particular there is $m\le n$ and $\gamma\in b_m$ such that $\tau_\gamma(N_m[s,s'])$ diverges.\\
  Since $N_m\geob y_m$ and $\tau_\gamma(N_m[s,s'])\Ua$, the co-induction gives $(\gamma_k)_k$ such that $\gamma_0=\gamma$ and for\linebreak all $k$, $\gamma_k=c_1^k\cons\cdots c_{g_{N_m}(k)}^k\cons\gamma_k'$ and $\gamma_{k+1}\in\bigcup_{i\le g_{N_m}(k)}a_i^k$. In this case we can define $(\alpha_k)_k$ as follows:
    \begin{align*}
      \alpha_0 &=\alpha  &
      \forall k,\ \alpha_{k+1} &= \gamma_k
    \end{align*}
    This is sufficient since:
    \begin{align*}
      m &\le n = g_N(0) &
      g_{N_m}(k) &\le sup_{j\le n}g_{N_j}(k) = g_N(k+1)
    \end{align*}
\end{proof}

\subsubsection{Inequational completeness}\label{sssec:red}
  
\begin{theorem}[Inequational full completeness]\label{th:FAtecn} For all $M,N\in\Lamb$, 
  $$M\leob N \quad\quad\Rta\quad\quad \llb M\rrb^{\vec x}\subseteq \llb N\rrb^{\vec x}.$$
\end{theorem}
\begin{proof}
  We will prove the equivalent (by Theorem~\ref{th:caracFAwT}) statement:
  \begin{itemize}
  \item[] Let $\alpha\in D$ and $a_0,\dots,a_k\in\mathcal{A}_f(D)$.\\
    Let $\{x_0,\dots,x_k\}\supseteq \mathrm{FV}(M)$ be a set of variables, and let $[s]=[\bareps_{a_0}/x_0\cdots\bareps_{a_k}/x_k]$.\\
    If\footnote{Recall that $M\Da_{\!\! n}$ means that $M$ may-head converges in at most $n$ steps} $\tau_\alpha(M[s])\Da_{\!\! n}$ and $\tau_\alpha(N[s])\Ua$ then $M \not\leob N$.
  \end{itemize}
  The statement is proved by induction on the length $n$ of the reduction $\tau_\alpha(M[s])\Da_{\!\! n}$:
  \begin{itemize}
  \item The case $n=0$:\\
    Then $\tau_\alpha(M[s])$ is in normal form without free variables, which is impossible.
  \item The case $n\ge 1$:\\
    Since $\tau_\alpha(M[s])\Da_{\!\! n}$, by applying the sensibility for \Lam{D}, the interpretation of $\tau_\alpha(M[s])\Da_{\!\! n}$ is non empty. By Remark~\ref{rk:typ_sub}, the interpretation of $M$ is also non empty. Thus, reapplying the sensibility, $M$ is converging to a head-normal form $M\rta_h^*\lambda y_1...y_n.z\ M_1\cdots M_m$. We can then make some assumptions:
    \begin{itemize}
    \item We can assume that $N\rta_h^*\lambda y_1...y_{n'}.z'\ N_1\cdots N_{m'}$:\\ 
      In fact, if $N$ does not converge then trivially $M\not\leob N$.
    \item We can assume that $n'\ge n$:\\ 
      In fact, if $n'\mathrm{<}n$ then we can always define $N'=\lambda y_1...y_{n'}y_{n'+1}...y_n.z'\ N_1\cdots N_{m'}\ y_{n'+1}\cdots y_n$ (with $y_{n'+1}...y_n\not\in\FV(z'\ N_1\cdots N_{m'})$), and we would have $N'\equivob N$ and $\tau_{\alpha}(N'[s])\Ua$.
    \item We can assume that $n\mathrm{=}0$:\\ 
      In fact, let $a_0\cons\cdots a_n\cons\alpha'\mathrm{=}\alpha$, $[s']\mathrm{=}[\bareps_{a_0}/y_1,\dots,\bareps_{a_n}/y_n]$, $N'=\lambda y_{n+1}...y_{n'}.z'\ N_1\cdots N_{m'}$ and $M'=z\ M_1\cdots M_{m}$. Since $\tau_\alpha(M[s])\rta^*\tau_{\alpha'}(M'[s,s'])$ (resp. $\tau_\alpha(N[s])\rta^*\tau_{\alpha'}(N'[s,s'])$), by confluence and standardization theorems (Th.~\ref{th:confluence} and Th.\ref{th:standardisation}), the convergences of $\tau_\alpha(M[s])$ (resp. $\tau_\alpha(N[s])$) and  $\tau_{\alpha'}(M'[s,s'])$ (resp. $\tau_{\alpha'}(N'[s,s'])$) are equivalent. Applying Theorem~\ref{th:invHeadReducibility}, we thus have $\tau_{\alpha'}(M'[s,s'])\Da_{\!\! n}$ and $\tau_{\alpha'}(N'[s,s'])\Ua$.\\ Moreover $M'\leob N'\Lra M\leob N$ so that the property on $M'$ and $N'$ is equivalent to the same property on $M$ and $N$.
    \item We can assume that $z'=z=x_0$:\\
      Since $\{x_0\dots x_k\}\supseteq \mathrm{FV}(M)$, there is $j\le k$ such that $z=x_j$, for simplicity we assume that $j=0$. Then we can remark that by Item \eqref{eq:y=y'} of Lemma~\ref{lemma:BTlemmes}, either $M\not\leob N$ or~$z'=z=x_0$, we will thus continue with the second case.
    \end{itemize}
    Altogether we have:
    \begin{align*}
      M &\rta_h^*x_0\ M_1\cdots M_m & \ \quad N&\rta_h^*\lambda y_1...y_{n'}.x_0\ N_1\cdots N_{m'}
    \end{align*}
    The case $M=x_0$ corresponds exactly to the hypothesis of Lemma~\ref{lemma:fonda} that concludes \linebreak by $M=x_0\not\leob N$. We are now assuming that $m\ge 1$.

    By Lemma~\ref{lemma:BTlemmes}, either $M\not\leob N$ or the following holds:
    \begin{itemize}
      \item $m=m'-n'$, and in particular $m\le m'$
      \item for $i\le m $,  $M_i\leob N_i$ 
      \item for $m<i\le m'$, $y_{i-m}\leob N_i$.
    \end{itemize}
    We will assume that $m=m'-n'$ and then refute $M_i\leob N_i$  or $y_{i}\leob N_{m+i}$ for some $i\le n'$; we then conclude that $M\not\leob N$.
    
    In the following we unfold 
    \begin{itemize}
    \item $\alpha=b_1\cons\cdots\cons b_{n'}\cons\alpha'$,
    \item $a_0=\{\beta_0\dots\beta_r\}$,
    \item for all $t\le r$, $\beta_{t}=c^{t}_1\cons\cdots c^{t}_{m}\cons\beta'_{t}$,
    \item and  for all $t\le r$, $\beta'_{t}=c^{t}_{m+1}\cons\cdots c^{t}_{m'}\cons\beta''_{t}$.
    \end{itemize}
    Moreover we set $[s']=[\bareps_{b_1}/y_1\dots \bareps_{b_{n'}}/y_{n'}]$.\\

    Then we have:
    \begin{eqnarray}
      \tau_\alpha(M[s]) & {\rta^*} & \tau_\alpha(\bareps_{a_0}\ M_1[s]\cdots M_m[s]) \\
      & \ruledarrow{\bar\tau}{h}{m}\ruledarrow{\tau\bar\tau}{h}{} & \Sigma_{\{t\le r\mid \alpha\le\beta'_t\}}\Pi_{i\le m}\Pi_{\gamma\in c^{t}_i}\tau_\gamma(M_i[s]). \label{eq:FAtecnIntern}
    \end{eqnarray}
    By Theorem~\ref{th:invHeadReducibility}, $\tau_\alpha(\bareps_{a_0}\ M_1[s]\cdots M_m[s])\Da_{\!\! n}$. Moreover, since the head reduction \eqref{eq:FAtecnIntern} is prefix of any head reduction sequence starting from $\tau_\alpha(\bareps_{a_0}\ M_1[s]\cdots M_m[s])$, the \linebreak test~$\Sigma_{\{t\le r\mid \alpha\le\beta'_t\}}\Pi_{i\le m}\Pi_{\gamma\in c^{t}_i}\tau_\gamma(M_i[s])$ head converges in $(n-m-1)$ steps so that there exists~$t_0\mathrm{\le}r$ such that $\alpha\le\beta'_{t_0}$ and for all $i\le m$ and all $\gamma\in c^{t_0}_i$, we have $M_i[s]\Da_{n-1}$.

    Similarly we have: 
    \begin{eqnarray*}
      \tau_\alpha(N[s])  & {\rta^*} & \tau_{\alpha}(\lambda y_1...y_{n'}.\bareps_{a_0}\ N_1[s]\cdots N_{m'}[s]) \\
      & \ruledarrow{\tau}{}{n'} & \quad \tau_{\alpha'}(\bareps_{a_0}\ N_1[s,s']\cdots N_{m'}[s,s'])) \\
      & \ruledarrow{\bar\tau}{}{m'} & \quad \tau_{\alpha'}(\Sigma_{t\le r}\bar\tau_{\beta''_{t}}(\Pi_{i\le m'}\Pi_{\gamma\in c^{t}_i}\tau_\gamma(N_i[s,s']))) \\
      & \ruledarrow{\tau\bar\tau}{}{} & \quad \Sigma_{t\le r\mid \alpha'\le \beta''_t}\Pi_{i\le m'}\Pi_{\gamma\in c^{t}_i}\tau_\gamma(N_i[s,s']).
    \end{eqnarray*}
    Thus, by standardization (Th.~\ref{th:standardisation}), $\Sigma_{t\le r\mid \alpha'\le \beta''_t}\Pi_{i\le m'}\Pi_{\gamma\in c^{t}_i}\tau_\gamma(N_i[s,s'])$ diverges. Thus there are two cases:
    \begin{itemize}
    \item Either $\alpha'\not\le\beta''_{t_0}$: which is impossible since $\alpha\mathrm{\le}\beta'_{t_0}$.
    \item Or there are $i\le m'$ and $\gamma\in c_i^{t_0}$ such that $\tau_\gamma(N_i[s,s'])$ diverges. 
      \begin{itemize}
      \item Either $i\le m$:\\
        Then since $\tau_\gamma(M_i[s,s'])=\tau_\gamma(M_i[s])\Da_{n-1}$, the induction hypothesis yields \linebreak that $M_i\not\leob N_i$.
      \item Or $m<i$:\\
        Since $\alpha\le\beta'_{t_0}$ we have $b_{i-m}\ge c^{t_0}_i$ and $\gamma\le\gamma'\in b_{i-m}$. Moreover, using Theorem~\ref{th:caracFAwT} and $\gamma\le\gamma'$, we have that $\tau_{\gamma'}(N_i[s,s'])$ diverges. Thus we can apply Lemma~\ref{lemma:fonda} that results in $y_{i-m}\not\leob N_i$.
      \end{itemize}
    \end{itemize}
  \end{itemize}
\end{proof}

\begin{theorem}[Hyperimmunity implies full abstraction]\label{th:FA}
  Any extensional \Kweb $D$ that is hyperimmune and sensible for $\Lam{D}$ is inequationally fully abstract for the pure $\lambda$-calculus.
\end{theorem}
\begin{proof}
  {\em Inequational adequacy:} inherited from the inequational sensibility of $D$ for $\Lam{D}$. Indeed, for any $M,N\in\Lamb$ and $C\in\Lcont$, if $\llb M\rrb_D^{\vec x}\subseteq \llb N\rrb^{\vec x}_D$ and if $C\llc M\rrc\Da$, 
then by sensi-\linebreak bility~$\llb C\llc N\rrc\rrb^{\vec x'}_D \supseteq \llb C\llc M\rrc\rrb^{\vec x'}_D\neq\emptyset$ and (still by sensibility) $\llb C\llc N\rrc\rrb^{\vec x'}_D$ converges.

  {\em Inequational completeness:} for all $M,N\mathrm{\in}\Lamb$ such that $\llb M\rrb^{\vec x\!}\not\subseteq\llb N\rrb^{\vec x\!}$, there is $(\vec a,\alpha)\in\llb M\rrb^{\vec x\!}\mathrm{-}\llb N\rrb^{\vec x\!\!}$, thus by Theorem~\ref{th:FAtecn}, $M\not\leob N$.
\end{proof}


  \subsection{Full abstraction implies hyperimmunity}
    \label{ssec:FAtoH}
    \label{ssec:FAtoH2}
%

\subsubsection{The counterexample}\ \newline
In this section, we are assuming that $D$ is sensible for \Lam{D} but is not hyperimmune. Then we will construct a counterexample $(\J_g\ \underline{0})$ for the full abstraction such that $(\J_g\ \underline{0})\equivob \I$ and $\llb\J_g\ \underline{0}\rrb\neq \llb\I\rrb$ resulting in Theorem~\ref{th:countex2}. 

 By Definition~\ref{def:hyperim}, if $D$ is hyperimmune, then there exist a recursive $g:(\Nat\rta\Nat)$ and a family~$(\alpha_n)_{n\ge 0}\in D^\Nat$ such that $\alpha_n=a_{n,1}\cons\cdots\cons a_{n,g(n)}\cons \alpha_n'$ with $\alpha_{n+1}\in \bigcup_{k\le g(n)}a_{n,k}$. 

We will use the function $g$ for defining a term $\J_g$ (Eq.~\ref{eq:defA2}) such that $(\J_g\ \underline{0})$ is observationally equal to the identity in $\Lamb$ (Lemma~\ref{lemma:JequivI}) but can be distinguished in $\Lam{D}$ (Cor.~\ref{cor:divergence2}). From this latter statement and the full abstraction for \Lam{D} (Th.~\ref{th:FAwT}), we will obtain that $\llb \J_g\ \underline{0}\rrb_{D}\neq\llb \Id\rrb_{D}$, and thus we conclude with Theorem~\ref{th:countex2}.


\medskip

\noindent
Let $(\G_n)_{n\in \mathbb{N}}$ be the sequence of closed $\lambda$-terms defined by: 
\begin{equation}
 \G_n:= \lambda ue x_1...x_{g(n)}.e\ (u\ x_1)\ \cdots\ (u\ x_{g(n)}) \label{eq:defG2}
\end{equation}

\noindent The recursivity of $g$ implies that of the sequence $\G_n$. We can thus use the Proposition~\ref{prop:autorecursivity} that build~$\G\in \Lamb$ such that:
\begin{equation}
 \G\ \underline{n}\rta^* \G_n. \label{eq:redG2}
\end{equation}

\noindent Recall that $\boldsymbol{S}$ denotes the Church successor function and $\boldsymbol{\Theta}$ the Turing fixedpoint combinator. We define:
\begin{equation}
 \J_g :=\boldsymbol{\Theta}\ (\lambda uv. \G\ v\ (u\ (\boldsymbol{S}\ v))). \label{eq:defA2}
\end{equation}
Then: 
\begin{equation}
 \J_g\ \underline{n} \rta^* \G_n\ (\J_g\ \underline{n\+1}). \label{eq:redA2}
\end{equation}

\begin{lemma}\label{lm:cefact}
  For all $n\in \Nat$, all $\alpha\in D$ and all $b=\{\beta_1,....,\beta_k\}\subseteq D$, let:
  \begin{itemize}
  \item $\alpha=a_1\cons\cdots\cons a_{g(n)}\cons\alpha'$,
  \item for all $j\le k$, $\beta_j=b_{j,1}\cons\cdots\cons b_{j,g(n)}\cons\beta_j'$,
  \end{itemize}
  we have:
  $$\tau_{\alpha}(\J_g\ \underline n\ \bareps_b)\ \rta^*\rta_h\ \Sigma_{\{j\le k\mid \alpha'\le\beta'_j\}}\Pi_{i\le g(n)}\Pi_{\gamma\in b_{j,i}}\tau_\gamma(\J_g\ \underline{n\+ 1}\ \bareps_{a_i}).$$
\end{lemma}
\begin{proof}
  We can reduce:
  \begin{eqnarray*}
    \tau_{\alpha}(\J_g\ \underline n\ \bareps_b)\hspace{-0.5em} 
    &\Ruledarrow{Eq\eqref{eq:redA2}}{}{*} & \tau_{\alpha}(\G\ \underline n\ (\J_g\ \underline{n\+ 1})\ \bareps_b )\\
    &\Ruledarrow{Eq\eqref{eq:defG2}}{}{*} & \tau_{\alpha}(\G_n\ (\J_g\ \underline{n\+ 1})\ \ \bareps_b)\\
    &\Ruledarrow{Eq\eqref{eq:redG2}}{}{*} & \tau_{\alpha}\Bigl((\lambda ue\vec x^{g(n)}.e\ (u\ x_1)\ \cdots\ (u\ x_{g(n)}))\ (\J_g\ \underline{n\+ 1})\ \ \bareps_b\Bigr)\\
    &\ruledarrow{\beta}{h}{2}& \tau_{\alpha}\Bigl(\lambda\vec x^{g(n)}.\bareps_b\ (\J_g\ \underline{n\+ 1}\ x_1)\ \cdots\ (\J_g\ \underline{n\+ 1}\ x_{g(n)})\Bigr)\\
    &\ruledarrow{\tau}{h}{g(n)}& \tau_{\alpha'}\Bigl(\bareps_b\ (\J_g\ \underline{n\+ 1}\ \bareps_{a_1})\ \cdots\ (\J_g\ \underline{n\+ 1}\ \bareps_{a_{g(n)}})\Bigr)\\
    &\ruledarrow{\bar\tau}{h}{g(n)}& \tau_{\alpha'}(\Sigma_{j\le k}\bar\tau_{\beta_j'}(\Pi_{i\le g(n)}\Pi_{\gamma\in b_{j,i}}\tau_\gamma(\J_g\ \underline{n\+ 1}\ \bareps_i)))\\
    &\ruledarrow{\tau\bar\tau}{h}{}& \Sigma_{\{j\le k\mid \alpha'\le\beta'_j\}}\Pi_{i\le g(n)}\Pi_{\gamma\in b_{j,i}}\tau_\gamma(\J_g\ \underline{n\+ 1}\ \bareps_{a_i})
  \end{eqnarray*}
\end{proof}

\begin{lemma}\label{lemma:JequivI}
  For all $n$, we have $\J_g\ \underline n \equivob \I$.
\end{lemma}
\begin{proof}
  Let \Dinf be defined as in Example~\ref{example:1}, it is fully abstract for \Hst.\footnote{Notice that the full abstraction of \Dinf for \Hst, that has been proved for decade \cite{Hyl75,Wad76}, can be recovered as we  have seen in Example~\ref{ex:hyperimmunity} that \Dinf is hyperimmune.} It results that it is sufficient to verify that $\llb \J_g\ \underline n\rrb_{\Dinf} = \llb \I \rrb_{\Dinf}$, or equivalently (Th.~\ref{th:FA}) to verify that :
  $$ \forall \alpha\in\Dinf, \tau_\alpha(\J_g\ \underline n)\Da \Lra \tau_\alpha(\I)\Da.$$
  Trivially $\tau_{a_0\cons\alpha}(\I)$ converges iff there is $\beta$ such that $\alpha\le \beta\in a_0$. Conversely we can prove by induction on $a_0$ that $\tau_\alpha(\J_g\ \underline n\ \bareps_{a_0})$ converges iff there is $\beta$ such that $\alpha\le \beta\in a_0$ and conclude by extensionality.

  If we denote $\alpha= a_1\cons\cdots\cons a_{g(n)}\cons\alpha'$, Lemma~\ref{lm:cefact} gives that:
    \begin{eqnarray*}
      \tau_\alpha(\J_g\ \underline n\ \bareps_{a_0}) &{\rta^*\rta_h}& \Sigma_{\{b_1\cons\cdots b_{g(n)}\cons\beta'\in a_0\mid \alpha'\le\beta'\}}\Pi_{i\le g(n)}\Pi_{\gamma\in b_i}\tau_\gamma(\J_g\ \underline{n\+1}\ \bareps_{a_i}). \\
    \end{eqnarray*}
    By induction hypothesis and standardisation, this test converges iff there is $\beta=b_{1\!}\cons\cdots b_{g(n)\!}\cons\beta'\!\in a_0$ such that  $\alpha'\le \beta'$ and for all $i\le g(n)$ and all $\gamma\in b_i$, $\gamma\le \delta\in a_i$, {\em i.e.}, for all $i$, $b_i\le a_i$. Equivalently, this test converges iff $\alpha\le \beta\in a_0$. Thus, using the standardization (Th.~\ref{th:standardisation}), $\tau_\alpha(\J_g\ \underline n\ \bareps_{a_0})$ converges iff $\alpha\le \beta\in a_0$.
\end{proof}

\begin{lemma}\label{lemma:ce}
  For all $n\in \Nat$, all $\alpha\in D$ and all $b\in\Achf D$, if $\beta\not\ge\alpha$ for all $\beta\in b$, then:
  $$\tau_{\alpha}(\J_g\ \underline n\ \bareps_b) \Ua$$
\end{lemma}
\begin{proof}
  Let $\{\beta_1,....,\beta_l\}=b$ and, for all $j\le l$, let $b_{j,1}\cons\cdots\cons b_{j,l}\cons\beta_j'=\beta_j$.\\
  We are proving by induction on $k$ that there is no convergence in $k$ steps:\footnote{\label{note:co-induction}We could have used a co-induction, but justifying the productivity is not easy (it uses Theorem~\ref{th:invHeadReducibility}).}\todo{Change in the thesis?}\\ 
  We assume that $\tau_{\alpha}(\J_g\ \underline n\ \bareps_b)\Da_{\!\! k\+1}$.\\
  From Lemma~\ref{lm:cefact}, we have: 
  $$\tau_{\alpha}(\J_g\ \underline n\ \bareps_b)\ \rta^*\rta_h\ \Sigma_{\{j\le l\mid \beta'_j\le\alpha'\}}\Pi_{i\le g(n)}\Pi_{\gamma\in b_{j,i}}\tau_\gamma(\J_g\ \underline{n\+ 1}\ \bareps_{a_i})$$
  By Theorem~\ref{th:invHeadReducibility}, and since the last head reduction was necessary, the resulting term converges in $k$ steps. Thus there exists $j\le l$ such that $\beta_j'\ge\alpha'$ and for all $i\le g(n)$ and each $\gamma\in b_{j,i}$, $\tau_\gamma(\J_g\ \underline{n\+ 1}\ \bareps_{a_i})$ converges in $k$ steps.\\
  Let $j\le l$ be such that $\beta_j'\ge\alpha'$. Since $\beta_j\not\ge\alpha$, there is $i$ such that $b_{j,i}\not\le a_i$, {\em i.e.}, there is $\gamma\in b_{j,i}$ such that for all $\delta\in a_i$, $\gamma\not\le \delta$ and by induction we get a contradiction to $\tau_\gamma(\J_g\ \underline{n\+ 1}\ \bareps_{a_i})\Da_{\!\! k}$.
\end{proof}

We recall that $(\alpha_n)_n$ is given by the counterexample of the hyperimmunity, and that for \linebreak all~$n$,~$\alpha_n=a_{n,1}\cons\cdots\cons a_{n,g(n)}\cons \alpha_n'$ and $\alpha_{n+1}\in \bigcup_{k\le g(n)}a_{n,k}$.

\begin{lemma}\label{lemma:divdemo}
  For any $n\in \Nat$ and any anti-chain $b=\{\alpha_n,\beta_1,....,\beta_k\}$, then:
  $$\tau_{\alpha_n}((\J_g\ \underline n)\ \bareps_b) \Ua.$$
  In particular, $\tau_{\alpha_0}(\J_g\ \underline 0\ \bareps_{\alpha_0}) \Ua$.
\end{lemma}
\begin{proof}
  We unfold $\beta_j=b_{j,1}\cons\cdots\cons b_{j,g(n)}\cons\beta'_j$.\\
  We are proving by induction on $k$ that there is no convergence in~$k$ steps:\footnote{See footnote~\ref{note:co-induction}}\todo{Change in the thesis?}\\ 
  We assume that $\tau_{\alpha}(\J_g\ \underline n\ \bareps_b)\Da_{\!\! k\+1}$.\\
  From Lemma~\ref{lm:cefact}, we have: 
  \begin{eqnarray*}
    \tau_{\alpha}(\J_g\ \underline n\ \bareps_b)\ & {\rta^*\rta_h}\ &  \Pi_{i\le g(n)}\Pi_{\gamma\in a_{ni}}\tau_\gamma(\J_g\ \underline{n\+ 1}\ \bareps_{a_i})\quad  +\quad \Sigma_{\{j\le l\mid \alpha'_n\le\beta'_j\}}\Pi_{i\le g(n)}\Pi_{\gamma\in b_{ji}}\tau_\gamma(\J_g\ \underline{n\+ 1}\ \bareps_{a_i}).
  \end{eqnarray*}
  By Theorem~\ref{th:invHeadReducibility}, and since the last head reduction was necessary, the resulting term converges in~$k$ steps. Thus one of the addends should converges in~$k$ steps, however:
  \begin{itemize}
  \item The fist member $\Pi_{i\le g(n)}\Pi_{\gamma\in a_{ni}}\tau_\gamma(\J_g\ \underline{n\+ 1}\ \bareps_{a_i})$ does not since there is $i\le g(n)$ such that $\alpha_{n\+1}\in a_{ni}$ and by induction, $\tau_{\alpha_{n\+1}}(\J_g\ \underline{n\+ 1}\ \bareps_{a_i})$ cannot converges in $k$ steps. 
  \item The second member of the sum diverges by Lemma~\ref{lemma:ce}.\\
   For any $j\le l$ such that $\beta_j'\ge\alpha_n'$  we know that $\beta_j\not\ge\alpha_n$ since $\{\alpha_n,\beta_1,...,\beta_l\}$ is an anti-chain. Thus there is always $i\le g(n)$ such that $b_{j,i}\not\le a_{n,i}$, {\em i.e.}, there is $\gamma\in b_{j,i}$ such that for \linebreak all $\delta\in a_{n,i}$, $\gamma\not\le \delta$. We can conclude by Lemma~\ref{lemma:ce} that $\tau_\gamma(\J_g\ \underline{n\+ 1}\ \bareps_{a_i})$ diverges.
  \end{itemize}
\end{proof}

\begin{theorem}[Full abstraction implies Hyperimmunity]\label{th:countex2}
  If $D$ is not hyperimmune, but sensible for $\Lam{D}$, then it is not fully abstract for the $\lambda$-calculus.
\end{theorem}
\begin{proof}
  Since $\tau_{\alpha_0}(\boldsymbol{I}\ \bareps_{\alpha_0})\stackrel{\beta}{\rta}_h\stackrel{\tau\bar\tau}{\rta}_h\epsilon$, we have that~$\llb \tau_{\alpha_0}(\Id\ \bareps_{\alpha_0})\rrb\mathrm{\neq}\emptyset$, while by Lemma~\ref{lemma:divdemo} we have that~$\llb\tau_{\alpha_0}(\J_g\ \underline{0}\ \bareps_{\alpha_0})\rrb\mathrm{=}\emptyset$, and thus $\llb\J_g\ \underline{0}\rrb\neq\llb\boldsymbol{I}\rrb$. Hence, by Lemma~\ref{lemma:JequivI}, $D$ is not fully abstract. 
\end{proof}

%
%
%
%
%
%
%
%


\bibliographystyle{plain}
\bibliography{article}

\appendix

\section{Appendix}

\subsection{Lemma \ref{lemma:test}} \ \newline

\begin{oneshot}{Lemma \ref{lemma:test}}
  If $D$ is sensible for $\Lamb_{\tau(D)}$ then:
  $$(\vec a b,\alpha)\in\llb M\rrb^{\vec y x}\ \ \Lra\ \ \ (\vec a,\alpha)\in\llb M[\bareps_b/x]\rrb^{\vec y},$$
  \vspace{-1em}
  $$(\vec a, \alpha)\in\llb M\rrb^{\vec y}\ \ \Lra\ \ \vec a\in\llb \tau_\alpha(M)\rrb^{\vec y}.$$
\end{oneshot}
\begin{proof}
For this proof we use the intersection type system of Figure~\ref{fig:tyTests}. Such a change of viewpoint replaces the statement by:
  $$\Gamma,x:a\vdash M:\alpha\ \ \Lra\ \ \Gamma\vdash M[\bareps_a/x]:\alpha$$
  $$\Gamma\vdash M:\alpha\ \ \Lra\ \ \Gamma\vdash \tau_\alpha(M)$$
\begin{itemize}
\item $\Gamma,x:a\vdash M:\alpha\ \ \Rta\ \ \Gamma\vdash M[\bareps_a/x]:\alpha\quad$ and $\quad\Gamma,x:a\vdash Q\ \ \Rta\ \ \Gamma\vdash Q[\bareps_a/x]$:\\
  By structural induction on $M$ and $Q$:
  \begin{itemize}
  \item If $M=x$: then $\alpha\le \beta\in a$ and by definition $\Gamma\vdash \bareps_a:\alpha$.
  \item If $M=y\neq x$: trivial.
  \item If $M= \lambda y. N$: then $\alpha=b\cons\beta$ and $\Gamma,y:b,x:a\vdash N:\beta$ thus by IH, $\Gamma,y:b\vdash N[\bareps_a/x]:\beta$ and thus $\Gamma\vdash M[\bareps_a/x]:\alpha$.
  \item If $M= N_1\ N_2$: then there exists $b$ such that $\Gamma,x:a\vdash N_1:b\cons\alpha$ and for all $\beta\in b$, $\Gamma,x:a\vdash N_2:\beta$. Thus by IH, $\Gamma\vdash N_1[\bareps_a/x]:b\cons\alpha$ and for all $\beta\in b$, $\Gamma\vdash N_2[\bareps_a/x]:\beta$ and thus $\Gamma\vdash M[\bareps_a/x]:\alpha$.
  \item If $M= \Sigma_i\bar\tau_{\alpha_i}(Q_i)$: then there exists $i$ such that $\alpha=\alpha_i$ and $\Gamma,x:a\vdash Q_i$. Thus by IH, $\Gamma\vdash Q_i[\bareps_a/x]$ and thus $\Gamma\vdash M[\bareps_a/x]:\alpha$.
  \item If $Q= \Sigma_iQ_i$: then there exists $i$ such that $\Gamma,x:a\vdash Q_i$. Thus by IH, $\Gamma\vdash Q_i[\bareps_a/x]$ and thus $\Gamma\vdash Q[\bareps_a/x]$.
  \item If $Q= \Pi_iQ_i$: then for all $i$, $\Gamma,x:a\vdash Q_i$. Thus by IH, for all $i$, $\Gamma\vdash Q_i[\bareps_a/x]$ and thus $\Gamma\vdash Q[\bareps_a/x]$.
  \item If $Q= \tau_\beta(M)$: then $\Gamma,x:a\vdash M:\beta$. Thus by IH, $\Gamma\vdash M[\bareps_a/x]:\beta$ and thus $\Gamma\vdash Q[\bareps_a/x]$.
  \end{itemize}
\item $\Gamma,x:a\vdash M:\alpha\ \ \Lra\ \ \Gamma\vdash M[\bareps_a/x]:\alpha$:\\
  and $\Gamma,x:a\vdash Q\ \ \Lra\ \ \Gamma\vdash Q[\bareps_a/x]$:\\
  By structural induction on $M$ and $Q$:
  \begin{itemize}
  \item If $M=x$ then $\Gamma\vdash \bareps_a:\alpha$ and by definition $\Gamma,x:a\vdash x:\alpha$, {\em i.e}, $\Gamma,x:a\vdash M:\alpha$
  \item If $M=y\neq x$: trivial.
  \item If $M= \lambda y. N$: then $\alpha=i_D(b\cons\beta)$ and $\Gamma,y:b\vdash N[\bareps_a/x]:\beta$ thus by IH, $\Gamma,y:b,x:a\vdash N:\beta$ and thus $\Gamma,x:a\vdash M:\alpha$.
  \item If $M= N_1\ N_2$: then there exists $b$ such that $\Gamma\vdash N_1[\bareps_a/x]:b\cons\alpha$ and for all $\beta\in b$, $\Gamma\vdash N_2[\bareps_a/x]:\beta$. Thus by IH, $\Gamma,x:a\vdash N_1:b\cons\alpha$ and for all $\beta\in b$, $\Gamma,x:a\vdash N_2:\beta$ and thus $\Gamma,x:a\vdash M:\alpha$.
  \item If $M= \Sigma_i\bar\tau_{\alpha_i}(Q_i)$: then there exists $i$ such that $\alpha=\alpha_i$ and $\Gamma\vdash Q_i[\bareps_a/x]$. Thus by IH, $\Gamma,x:a\vdash Q_i$ and thus $\Gamma,x:a\vdash M:\alpha$.
  \item If $Q= \Sigma_iQ_i$: then there exists $i$ such that $\Gamma\vdash Q_i[\bareps_a/x]$. Thus by IH, $\Gamma,x:a\vdash Q_i$ and thus $\Gamma,x:a\vdash Q$.
  \item If $Q= \Pi_iQ_i$: then for all $i$, $\Gamma\vdash Q_i[\bareps_a/x]$. Thus by IH, for all $i$, $\Gamma,x:a\vdash Q_i$ and thus $\Gamma,x:a\vdash Q$.
  \item If $Q= \tau_\beta(M)$: then $\Gamma\vdash M[\bareps_a/x]:\beta$. Thus by IH, $\Gamma,x:a\vdash M:\beta$ and thus $\Gamma,x:a\vdash Q$.
  \end{itemize}
\item $\Gamma\vdash \tau_\alpha(M)\ \ \Lra\ \ \Gamma\vdash M:\alpha$: by definition of the inference rule for $\tau_\alpha$
\end{itemize}
\end{proof}

\subsubsection{Lemma \ref{lemma:BTlemmes}}\ \newline

\begin{oneshot}{Lemma \ref{lemma:BTlemmes}}
  Let ${M=\lambda x_1...x_n.y\ M_1\cdots M_k}\in\Lamb$ and let $N=\lambda x_1...x_{n'}.y'\ N_1\cdots N_{k'}\in\Lamb$ be such that $M\leob N$. Then:
  \begin{enumerate}
  \item $y=y'$, \label{eq:y=y'}
  \item $n-k=n'-k'$, \label{eq:m-n}
  \item if $i\le k$ and $i\le k'$ then $M_i\leob N_i$, \label{eq:MeqN}
  \item if $i> k$ and $i\le k'$ then $x_{i-k}\leob  N_i$, \label{eq:Meqx}
  \item if $i\le k$ and $i> k'$ then $M_{i-k}\leob  x_i$. \label{eq:Meqx2}
  \end{enumerate}
\end{oneshot}
\begin{proof}

  In the following, $M=\lambda x_1...x_n.y\ M_1\cdots M_k$ and $N=\lambda x_1...x_{n'}.y'\ N_1\cdots N_{k'}$.

  If $y\neq y'$, then $M\not\leob N$, indeed:
  \begin{itemize}
  \item If $y'$ is free in $M$ and $N$ then by setting $C\llc.\rrc = (\lambda y'.\llc.\rrc)\ \Omega$ we have $C\llc M\rrc\Da$ and $C\llc N\rrc\Ua$.
  \item If $y'= x_j$ for $j\le n'$, then by setting $C\llc . \rrc = \llc.\rrc\ x_1\ \cdots\ x_{j-1}\ \Omega$ we have $C\llc M\rrc\Da$ and $C\llc N\rrc\Ua$.
  \end{itemize}

  Now we suppose that $M=\lambda x_1...x_n.y\ M_1\cdots M_k$ and $N=\lambda x_1...x_{n'}.y\ N_1\cdots N_{k'}$.

  If $n-k\neq n'-k'$, then $M\not\leob  N$:
  \begin{itemize}
  \item If $y$ is free in $M$ and $N$, then by setting $C\llc.\rrc = (\lambda y.\llc.\rrc\ x_1\cdots x_{n'+k})\ (\lambda z_1...z_{k'+k}u.u)\ \Om$ we have $C\llc M\rrc\Da$ and $C\llc N\rrc\Ua$:
  \item If $y=x_j$ for $j\le n'$, then by setting $C\llc.\rrc = \llc.\rrc\ x_1\cdots x_{j-1}\ (\lambda z_1...z_{k'+k}u.u)\ x_{i+1}\cdots x_{n'+k}\ \Om$ we have $C\llc M\rrc\Da$ and $C\llc N\rrc\Ua$.
  \end{itemize}

  Now we suppose that $n-k = n'-k'$.

  If there is $i$ such that $i\le k$, $i\le k'$ and $M_i\not\leob N_i$ then there is $C'\llc.\rrc$ such that $C'\llc M_i\rrc\Da$ and $C'\llc N_i\rrc\Ua$:
  \begin{itemize}
  \item If $y$ is free in $M$ and $N$, then by setting $C\llc.\rrc = (\lambda y.\llc.\rrc)\ (\lambda z_1...z_{k+k'}.C'\llc z_i\rrc)$ we have $C\llc M\rrc\Da$ and $C\llc N\rrc\Ua$.
  \item If $y=x_j$ for $j\le n'$, then by setting $C\llc.\rrc = \llc.\rrc\ x_1\cdots x_{j-1}\ (\lambda z_1...z_{k+k'}.C\llc z_i\rrc)$ we have $C\llc M\rrc\Da$ and $C\llc N\rrc\Ua$.
  \end{itemize}

  If there is $i$ such that $k<i\le k'$ and $x_{i-k}\not\leob N_i$ then there is $C'\llc.\rrc$ such that $C'\llc x_{i-k}\rrc\Da$ and $C'\llc N_i\rrc\Ua$:
  \begin{itemize}
  \item If $y$ is free in $M$ and $N$, then by setting $C\llc.\rrc = (\lambda y.\llc.\rrc\ x_1\cdots\ x_{n+k})\ (\lambda z_1...z_{k+k'}.C'\llc z_i\rrc)$ we have $C\llc M\rrc\Da$ and $C\llc N\rrc\Ua$.
  \item If $y=x_j$ for $j\le n'$, then by setting $C\llc.\rrc = \llc.\rrc\ x_1\cdots x_{j-1}\ (\lambda z_1...z_{k+k'}.C\llc z_i\rrc)\ x_{j+1}\cdots\ x_{n+k}$ we have $C\llc M\rrc\Da$ and $C\llc N\rrc\Ua$.
  \end{itemize}

  If there is $i$ such that $k'<i\le k$ and $M_i\not\leob x_{i-k'}$ then there is $C'\llc.\rrc$ such that $C'\llc M_i\rrc\Da$ and $C'\llc x_{i-k'}\rrc\Ua$:
  \begin{itemize}
  \item If $y$ is free in $M$ and $N$, then by setting $C\llc.\rrc = (\lambda y.\llc.\rrc\ x_1\cdots\ x_{n+k})\ (\lambda z_1...z_{k+k'}.C'\llc z_i\rrc)$ we have $C\llc M\rrc\Da$ and $C\llc N\rrc\Ua$.
  \item If $y=x_j$ for $j\le n'$, then by setting $C\llc.\rrc = \llc.\rrc\ x_1\cdots x_{j-1}\ (\lambda z_1...z_{k+k'}.C\llc z_i\rrc)\ x_{j+1}\cdots\ x_{n+k}$ we have $C\llc M\rrc\Da$ and $C\llc N\rrc\Ua$.
  \end{itemize}
\end{proof}

\end{document}